\documentclass[pre,twocolumn,superscriptaddress,preprintnumbers,amsmath,amssymb]{revtex4-1}
\usepackage{epsfig}
\usepackage{color}
\usepackage{amsmath}
\usepackage{amsfonts}
\usepackage{graphicx}
\usepackage{fancybox}
\usepackage{subfigure}
\usepackage{notes2bib}
\usepackage{fontenc}
\usepackage{amssymb}
\usepackage[version=3]{mhchem}
\usepackage{subfiles}

  \makeatletter
    \renewcommand\@make@capt@title[2]{%
     \@ifx@empty\float@link{\@firstofone}{\expandafter\href\expandafter{\float@link}}%
      {\textbf{#1}}\@caption@fignum@sep#2\quad}%
    \makeatother
   
\makeatletter 
\renewcommand{\fnum@figure}{\textbf{Figure~\thefigure}}
\makeatother

\begin{document}

%%%%%%%%%%%%%%%%%%%%%%%%%%%%%%
\bibliographystyle{pnas}

\title {Spontaneous Recovery of Superhydrophobicity on Nanotextured Surfaces}

%Author List
\author{Suruchi Prakash}
\affiliation{Department of Chemical \& Biomolecular Engineering, University of Pennsylvania, Philadelphia, PA 19104, USA}

\author{Erte Xi}
\affiliation{Department of Chemical \& Biomolecular Engineering, University of Pennsylvania, Philadelphia, PA 19104, USA}

\author{Amish J. Patel}
\email{amish.patel@seas.upenn.edu}
\affiliation{Department of Chemical \& Biomolecular Engineering, University of Pennsylvania, Philadelphia, PA 19104, USA}
%Date
\date{\today}

%%%%%%%%%%%%%%%%%%%%%%%%%%%%%%%%%%%%
\begin{abstract} % 245 as of now
%abstract of no more than 250 words
%Superhydrophobic surfaces, which are created by texturing hydrophobic surfaces, display large contact angles and low contact angle hysteresis, which confers them with numerous desirable properties such as self-cleaning and water repellency.
%
Rough or textured hydrophobic surfaces are dubbed superhydrophobic due to their numerous desirable properties, such as water repellency and interfacial slip.
%
%Superhydrophobic surfaces, which possess numerous desirable properties such as self-cleaning and water repellency, are created by texturing hydrophobic surfaces.
%, display large contact angles and low contact angle hysteresis, which confers them with .
%
Superhydrophobicity stems from an aversion for water to wet the surface texture, so that a water droplet in the superhydrophobic ``Cassie state'', contacts only the tips of the rough hydrophobic surface.
However, superhydrophobicity is remarkably fragile, and can break down due to the wetting of the surface texture to yield the ``Wenzel state'' under various conditions, such as elevated pressures or droplet impact.
Moreover, due to large energetic barriers that impede the reverse (dewetting) transition, this breakdown in superhydrophobicity is widely believed to be irreversible.
Using molecular simulations in conjunction with enhanced sampling techniques, here we show that on surfaces with nanoscale texture, water density fluctuations can lead to a reduction in the free energetic barriers to dewetting by circumventing the classical dewetting pathways.
In particular, the fluctuation-mediated dewetting pathway involves a number of transitions between distinct dewetted morphologies, with each transition lowering the resistance to dewetting.
Importantly, an understanding of the mechanistic pathways to dewetting and their dependence on pressure, allows us to augment the surface texture design, so that the barriers to dewetting are eliminated altogether and the Wenzel state becomes unstable at ambient conditions.
Such robust surfaces, which defy classical expectations and can spontaneously recover their superhydrophobicity, could have widespread importance, from underwater operation to phase change heat transfer applications.
\end{abstract}
%%%%%%%%%%%%%%%%%%%%%%%%%%%%%%%%%%%%

\maketitle
\raggedbottom

\section{Introduction}
Surface roughness or texture can transform hydrophobic surfaces into ``superhydrophobic'', and endow them with properties like water-repellency, self-cleaning, interfacial slip, and fouling resistance~\cite{Quere:2008cw,Nosonovsky:eu}. 
Each of these remarkable properties stems from the reluctance of water to penetrate the surface texture, so that a drop of water sits atop an air cushion in the so-called Cassie state, contacting only the top of the surface asperities.
However, water can readily penetrate the surface texture, yielding the Wenzel state~\cite{Lohse:PRL:2007,Butt:PNAS:2013} at elevated pressures~\cite{Lafuma:2003hc,Butt:PNAS:2013} or temperatures~\cite{Liu:2009uu}, upon droplet impact~\cite{koishi2009coexistence,Boreyko:2013fm}, as well as due to surface vibration~\cite{2014:Lei}, localized defects~\cite{Moulinet:2007aa}, or proximity to an electric field~\cite{Manukyan:2011jj}; superhydrophobicity is thus remarkably fragile and can break down due to the wetting of the surface texture under a wide variety of conditions.
%%%
%%%
To facilitate the recovery of superhydrophobicity and to afford reversible control over surface properties, significant efforts have focused on inducing the reverse Wenzel-to-Cassie dewetting transition.
However, a true Wenzel-to-Cassie transition has been elusive~\cite{Bormashenko-review}, with most reported instances making use of trapped air~\cite{Forsberg:2010ih,Manukyan:2011jj,Ras:PNAS:2012,Checco:PRL:2014} or generating a gas film using an external energy source~\cite{Krupenkin:2007aa,Liu:2011jh} to jumpstart the dewetting process.
Insights into why achieving a Wenzel-to-Cassie transition remains challenging are provided by macroscopic interfacial thermodynamics~\cite{Patankar:Langmuir:2004}, which suggests that the dewetting transition is impeded by a large free energetic barrier.
This ``classical'' barrier is attributed to the work of adhesion for nucleating a vapor-liquid interface at the base of the textured surface.
Consequently, the break down of superhydrophobicity upon wetting of the surface texture is widely believed to be irreversible~\cite{Lafuma:2003hc,Patankar:Langmuir:2004,Bormashenko-review}, so that once the texture wets, it remains in the wet state, even when the pressure is subsequently lowered or the electric field is switched off.
%%%

%%%
By employing atomistic simulations in conjunction with specialized sampling techniques, here we challenge this conventional wisdom, and uncover principles for the design of nanotextured surfaces, which can spontaneously recover their superhydrophobicity by dewetting their surface texture at ambient conditions.  
Our work builds upon recent theoretical and simulation studies, which have shown that water density fluctuations, which are not captured in macroscopic mean-field models, are enhanced at hydrophobic surfaces~\cite{LCW,LLCW,Mittal:PNAS:2008,Godawat:PNAS:2009,Patel:JPCB:2010}, and situate the interfacial waters at the edge of a dewetting transition~\cite{Patel:JPCB:2012}. 
Such enhanced fluctuations have also been shown to modulate the pathways to dewetting and lead to reduced dewetting barriers in several confinement contexts~\cite{tWC,maibaum2003coarse,Willard_tWC,Miller:PNAS:2007,Giacomello:PRL:2012,savoy2012molecular,shahraz2014kinetics,Remsing:PNAS:2015}. 
To investigate how fluctuations influence Cassie-Wenzel transitions on nanotextured surfaces, here we perform atomistic simulations of water adjacent to pillared surfaces, and employ the Indirect Umbrella Sampling (INDUS) method~\cite{Patel:JSP:2011} to characterize the free energetics of the transitions, the corresponding pathways, as well as their dependence on pressure. 
%wetting-dewetting transitions on these surfaces.
%
By comparing our results to macroscopic theory, we find that while water density fluctuations 
do not influence the pressure at which the Cassie-to-Wenzel wetting transition occurs, they are nevertheless crucial in the Wenzel-to-Cassie dewetting transition, that is, in the process of recovering superhydrophobicity when it breaks down.
In particular, fluctuations stabilize a non-classical dewetting pathway, which features crossovers between a number of distinct dewetted morphologies that precede the formation of the classical vapor-liquid interface at the basal surface; the non-classical pathway offers a lower resistance to dewetting, leading to reduced dewetting barriers. 
Importantly, by uncovering the nanoscale dewetting pathways, and in particular, by finding regions of the surface texture that are hardest to dewet, our results provide strategies for augmenting the surface texture to further destabilize the Wenzel state and reduce the barriers to dewetting.
On such rationally designed surfaces, the barriers to dewetting the texture can be eliminated altogether, so that the Wenzel state is no longer metastable, but has been rendered unstable at ambient conditions, and the superhydrophobic Cassie state can be spontaneously recovered from the wet Wenzel state.

%------------------------------------Fig 1--System -------------------------------------------------
\begin{figure}[ht]
     \begin{center}
     \vspace{-0.35in}
    \includegraphics[width=0.4\textwidth]{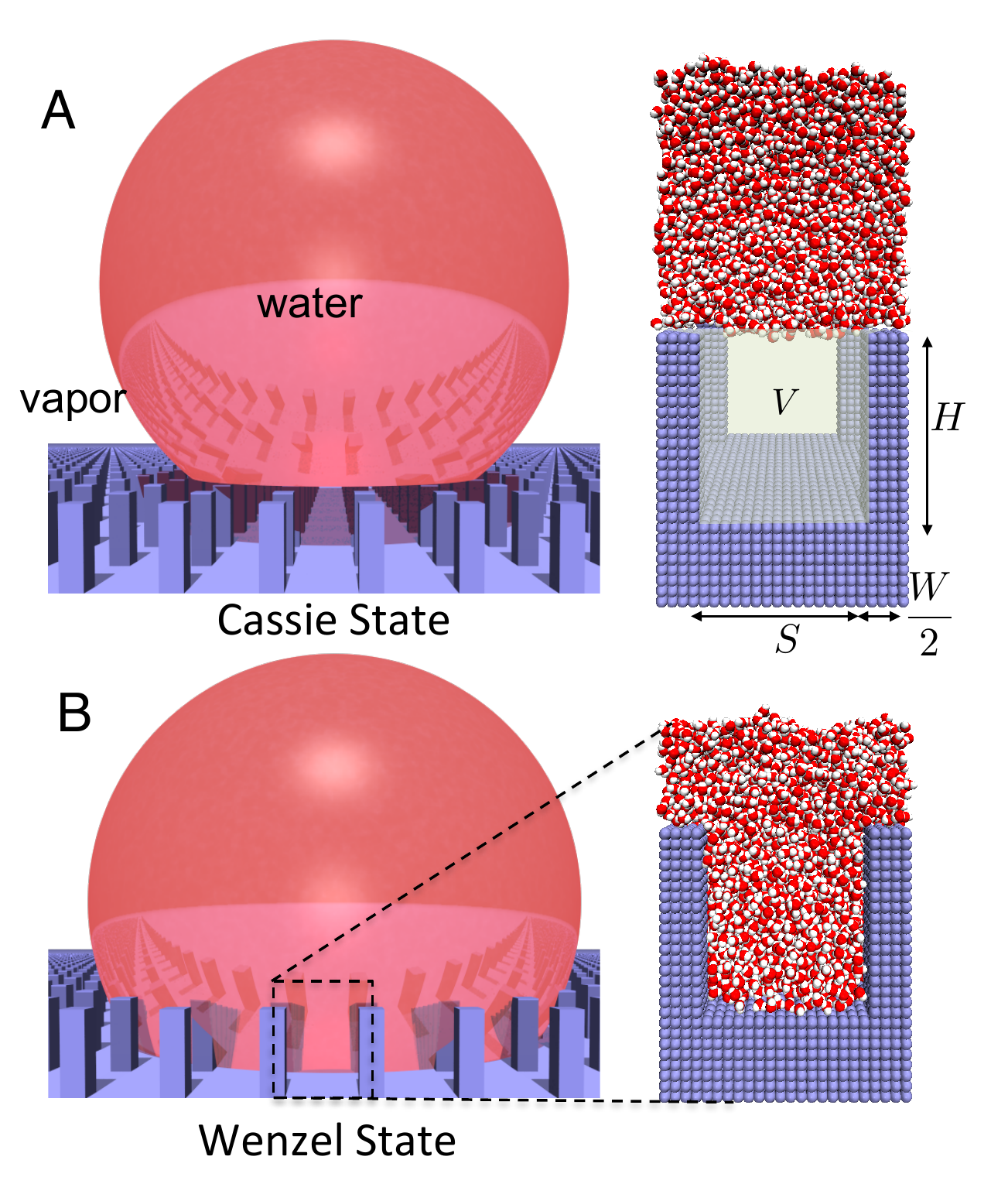}
   \end{center}
    \caption{ 
    \label{fig:model}
Water on textured hydrophobic surfaces can exist in either the Cassie or the Wenzel state.
(a) In the Cassie state, water is unable to penetrate the surface texture (blue) so that a water droplet (red) sits on a cushion of air, contacting only the top of the pillars.
As a result, there is minimal contact between water and the solid surface, leading to a small contact angle hysteresis and a large contact angle, which are critical in conferring superhydrophobicity to the surface.
Also shown is a simulation snapshot of the pillared surface that we study here (right), which consists of square pillars arranged on a square lattice, and is made of atoms (blue spheres) arranged on a cubic lattice.
The textured volume, $V$, as well as the dimensions that characterize the pillared nanotextured surface are highlighted; the width of the pillars is $W$=2~nm, their height is $H$=4.5~nm, and the inter-pillar spacing is $S$=4~nm.
 (b) In the Wenzel state, water wets the texture, so that there is extensive contact between water and the solid surface, leading to a large contact angle hysteresis and a smaller contact angle; in this state, the surface is no longer superhydrophobic.
 We define the normalized density, $\rhon$, in the textured volume to be the number of waters in $V$, normalized by the corresponding number of waters in the Wenzel state.
 }
\end{figure}
%--------------------------------------------------------------------------------------------------------------------
%
%\vspace{-0.4in}

%%%%%%%%%%%%%%%%%%%%%%%%%%%%%%%%%%%%
\section{Free Energetics of Cassie-Wenzel Transitions} 
%%%%%%%%%%%%%%%%%%%%%%%%%%%%%%%%%%%%

%%%
Figure~1 contrasts the behavior of water on textured surfaces in the Cassie and Wenzel states; water does not penetrate the surface texture in the Cassie state, but does so in the Wenzel state.
In Figure~1a (right), the particular textured surface morphology that we study here is shown, and consists of square pillars of height $H$ and width $W$, arranged on a square lattice, and separated by a distance $S$.
Also highlighted is the textured volume, $V$, which is devoid of water molecules in the Cassie state but is filled with waters in the Wenzel state.
The normalized water density, $\rhon$ in $V$, thus serves as a reliable order parameter to distinguish the Cassie and Wenzel states. 
In Figure~2a, we show the free energy, $\Delta F$, as a function of $\rhon$, obtained using molecular dynamics simulations in conjunction with Indirect Umbrella Sampling (INDUS)~\cite{Patel:JSP:2011}.
Here, $\rhon \equiv N/N_{\rm liq}$ with $N$ and $N_{\rm liq}$ being the number of water molecules in $V$, in the partially and fully wet states respectively.
Details pertaining to our simulation setups, the force-field parameters and algorithms employed, are included in Materials and Methods.
The simulated free energy profile, $\Delta F(\rhon)$, clearly shows two basins, Cassie at $\rhon\approx0$ and Wenzel at $\rhon\approx1$, which are separated by a large barrier. 
To uncover the importance of water density fluctuations on the free energetics of Cassie-Wenzel transitions, we first compare the simulated $\Delta F(\rhon)$ with classical expectations based on macroscopic interfacial thermodynamics, which does not account for fluctuations.
%
%%%

%%%
Macroscopic theory envisions dewetting proceeding through the nucleation of a vapor liquid interface perpendicular to the pillars, which then rises along the pillars.
The height of the interface above the base of the pillars is thus given by $h(\rhon) = H (1- \rhon)$, and the free energy, $\Delta F(\rhon)$, of a partially wet system relative to that in the Wenzel state is given by: 
\begin{equation}
\Delta F(\rhon) = \Delta F_{\rm barr} + [ \gamma\cos\theta A_{\rm side} + \Delta P V ](1 - \rhon),
    \label{eq:macro2}
\end{equation} 
where $\gamma$ is the vapor-liquid surface tension, $\theta$ is the water droplet contact angle on a flat surface, and $\Delta P$ is the difference between the system pressure and the co-existence pressure at the system temperature, $T$.
In addition, $\Delta F_{\rm barr} \equiv \gamma A_{\rm base} (1+\cos\theta)$ is always unfavorable (positive), and corresponds to the work of adhesion for creating the vapor-liquid interface, with $A_{\rm base} \equiv S ( S + 2W )$ being the basal area, and $A_{\rm side}=4WH$ being the area of the vertical faces of the pillars. 
Because $\cos\theta < 0$ for hydrophobic surfaces, the second term could be favorable (negative) if $\Delta P$ is sufficiently small, that is, if $\Delta P \le \Delta P_{\rm int} \equiv -\gamma\cos\theta A_{\rm side}/V$.
Thus, the two key features in the macroscopic theory $\Delta F(\rhon)$ are: (i) a large adhesion barrier at $\rhon\approx1$, which must be overcome to nucleate the vapor-liquid interface, and (ii) a linear portion of $\Delta F(\rhon)$ corresponding to the vapor-liquid interface rising along the pillars as $\rhon$ decreases.
A derivation of the above macroscopic theory expression and details of the comparison between simulation and theory are included in the Supporting Information. 
%%%

%%%
In particular, configurations with $0.2<\rhon<0.8$ (Figure~2c, Ia and Ib) are consistent with the macroscopic expectation, having dewetted regions with the vapor-liquid interface being perpendicular to the pillars, and at heights that are consistent with $h(\rhon) = H (1- \rhon)$; see Supporting Information. 
Theory also predicts that $\Delta F$ ought to vary linearly with $\rhon$; for $\Delta P=0$, the corresponding slope is expected to be $(-\gamma\cos\theta)4SW$, where $\gamma$ is the vapor-liquid surface tension and $\theta$ is the (flat surface) contact angle.
As seen in Figure~2a, the simulated $\Delta F(\rhon)$ is indeed linear for $0.2<\rhon<0.8$ (dashed line) with the fitted slope yielding a 
surface tension, $\gamma_{\rm fit} =67$~mJ/m$^2$, which agrees well with that of our water model, $\gamma_{\rm SPC/E} =63.6$~mJ/m$^2$~\cite{vega:2007}.
The behavior of the system for $0.2<\rhon<0.8$ is thus classical. 
Macroscopic theory also associates the barrier to dewetting with the work of adhesion to nucleate the vapor-liquid interface, which can be estimated as $\gamma_{\rm fit}(1+\cos\theta)S(S+2W)\approx 270$~$\kT$.
In contrast, the corresponding simulated barrier is only $118$~$\kT$ (Figure 2a).
Thus, while the ascension of the vapor-liquid interface along the pillars is classical, the nucleation of that vapor-liquid interface and the associated dewetting barriers, which are central to the recovery of superhydrophobicity, appear to be non-classical.
%
%%%

%----------------------------------FIGURE 2: Free energy------------------------------------------------------
\begin{figure*}[tb]
     \begin{center}
    \includegraphics[width=0.62\textwidth]{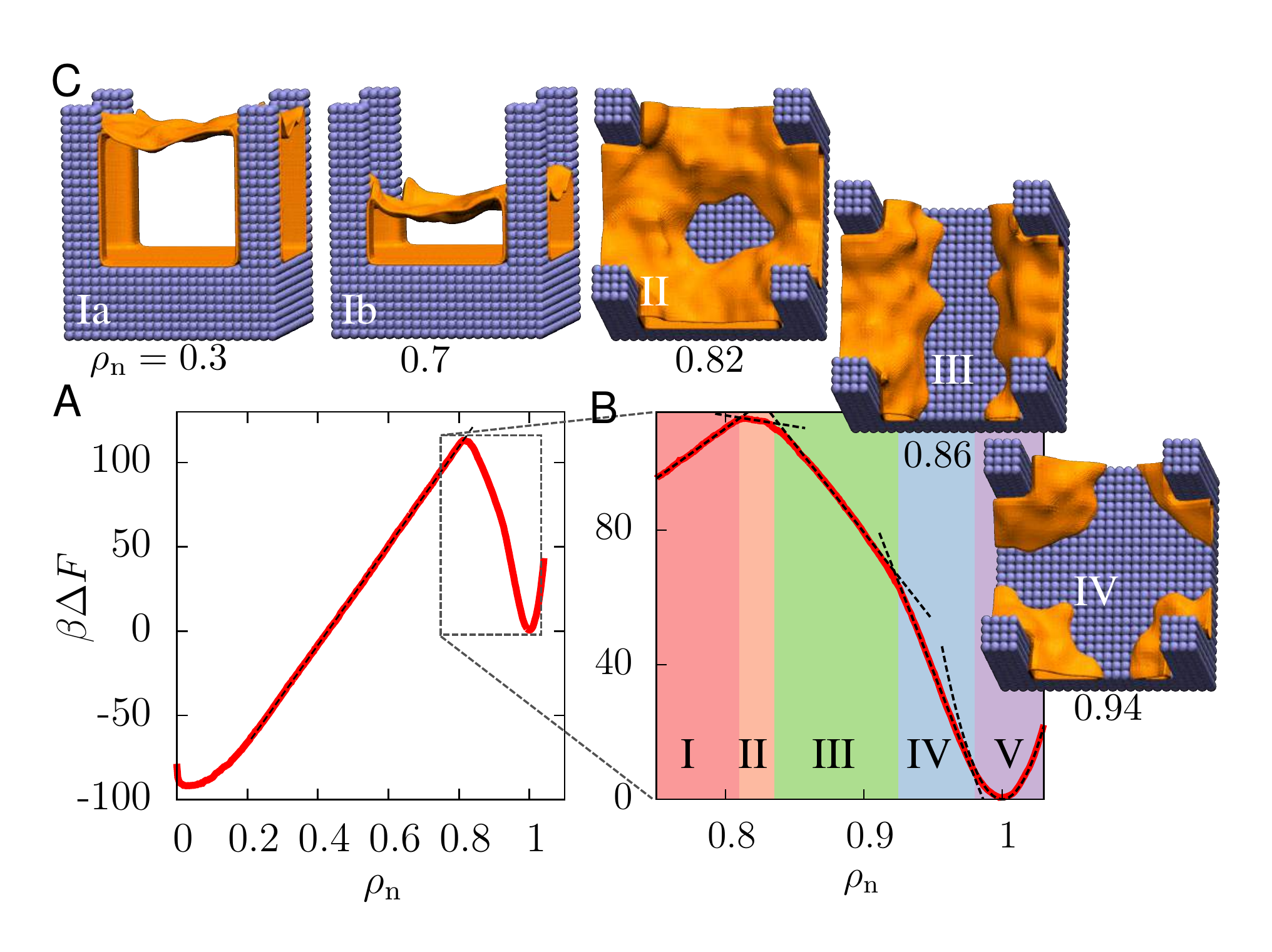}
     \end{center}
     \vspace{-0.2in}
    \caption{ 
     \label{fig:free}
Free energetics and pathways of wetting-dewetting transitions on a pillared surface.
(a) The simulated free energy, $\Delta F(\rhon)$ 
(in units of the thermal energy, $\kT\equiv\beta^{-1}$, with $k_{\rm B}$ being the Boltzmann constant and $T$ the temperature), features two basins that are separated by a large barrier.
For 0.2 $< \rhon <$ 0.8, $\Delta F$ varies linearly with $\rhon$, in agreement with macroscopic theory. 
However, the simulated barrier for dewetting ($118$~$\kT$) is found to be significantly smaller than the classical barrier ($270$~$\kT$). 
(b) Between the Wenzel state and the barrier ($0.75<\rhon<1$), $\Delta F(\rhon)$ is marked by several kinks, which demarcate five regions with distinct dewetted morphologies (dashed lines are a guide to the eye).
(c) Representative configurations corresponding to these regions are shown as interfaces encompassing the dewetted volumes (shown in orange, waters omitted for clarity).
Region V ($\rhon\approx1$) displays Gaussian fluctuations resulting in a parabolic basin.
Region IV ($0.93<\rhon<0.98$) is characterized by vapor pockets at the base of the pillars.
As $\rhon$ is reduced, vapor pockets grow, break symmetry, and merge to form a striped vapor layer between the pillars.
The stripe expands laterally in Region III ($0.83<\rhon<0.93$) until a nearly intact vapor-liquid interface is formed.
Region II ($0.81<\rhon<0.83$) is characterized by water molecules sticking to the center of the cell, and also contains the non-classical transition state (barrier), which eventually gives way to the classical Region I at $\rhon\approx0.8$.
\vspace{-0.1in}
 }
 \end{figure*}
%--------------------------------------------------------------------------------------------------------------------
% \vspace{-0.2in}

%%%%%%%%%%%%%%%%%%%%%%%%%%%%%%%%%%%%
\section{Fluctuations Facilitate Non-classical Dewetting Pathways With Reduced Barriers} 
%%%%%%%%%%%%%%%%%%%%%%%%%%%%%%%%%%%%
%%%
To understand why the dewetting barrier is significantly smaller than the classical expectation, we take a closer look at $\Delta F(\rhon>0.8)$ (Figure~2b) as well as the corresponding configurations (Figure~2c, II - IV), which are shown as representative instantaneous interfaces encompassing the dewetted regions (orange). 
A description of the algorithm used to compute the instantaneous interfaces, as well as the corresponding averaged interfaces are included in the Supporting Information. 
Interestingly, we observe a host of non-classical partially wet configurations preceding the formation of the classical vapor-liquid interface.
As $\rhon$ is decreased from 1, vapor pockets first form at the base of the pillars, then grow to round the corners around the pillars (Figure~2c, IV).
On further reducing $\rhon$, symmetry is broken as vapor pockets from opposite pillars merge to form stripes of vapor spanning the inter-pillar region (Figure~2c, III).
This change in the dewetted morphology coincides with a kink in $\Delta F(\rhon)$, suggesting that the system adopts a lower free energy path by transitioning from the vapor pocket to the stripe morphology.
Subsequent decrease in $\rhon$ results in another transition to a donut-shaped vapor layer (Figure~2c, II). 
$\Delta F(\rhon)$ displays a maximum in the donut morphology; the transition state thus corresponds to a configuration with water molecules sticking to the central region of the basal surface that is farthest from the pillars, rather than an intact vapor layer. 
Expelling the remaining water molecules to form an intact vapor layer is energetically favorable, as is the subsequent (classical) rise of the vapor-liquid interface along the pillars.
This novel and clearly non-classical pathway preceding the formation of a vapor-liquid interface, which is facilitated by nanoscopic water density fluctuations~\cite{Chandler:Nature:2005} and involves transitions between various dewetted configurations~\cite{Giacomello:PRL:2012,Remsing:PNAS:2015}, results in a smaller barrier for the Wenzel-to-Cassie transition than anticipated by macroscopic theory. 
%%%

%\vspace{-0.1in}
%%%%%%%%%%%%%%%%%%%%%%%%%%%%%%%%%%%%
\section{How Pressure Influences Barriers to Wetting and Dewetting} 
\begin{figure}[ht]
     \begin{center}
     \vspace{-0.5in}
    \includegraphics[width=0.5\textwidth]{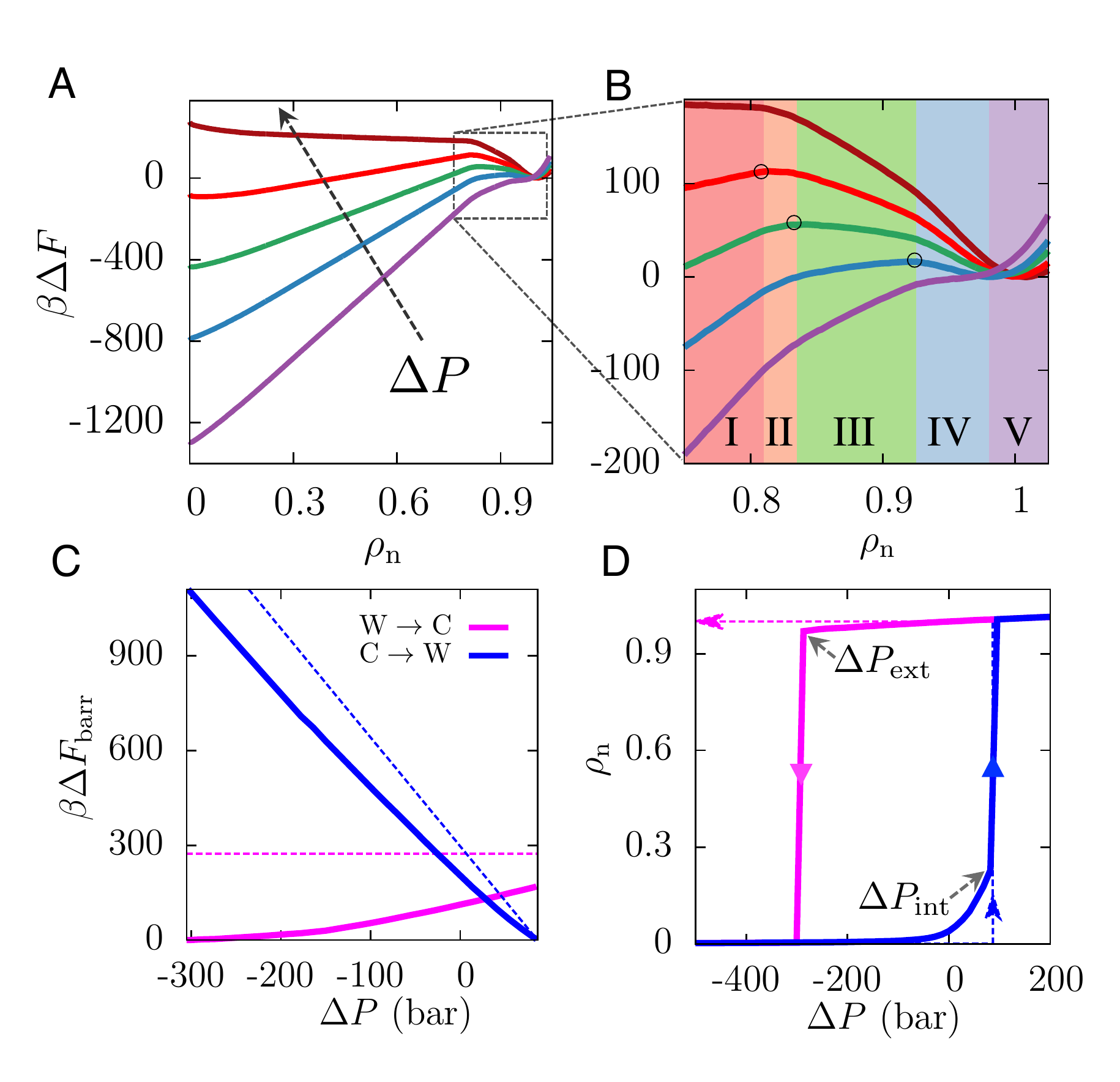}  
    \end{center}
   \vspace{-0.1in}
    \caption{ 
    \label{fig:pressure}
    Effect of pressure on Cassie-Wenzel transitions.
    (a) $\Delta F(\rhon)$ is shown for pressures ranging from -350 to 100 bar, with the arrow pointing in the direction of increasing pressure~(purple: -350 bar, blue: -200 bar, green: -100 bar, red: 0 bar, brown: 100 bar). 
    As pressure is increased, the slope of the classical region I decreases, destabilizing the Cassie state; conversely, as pressure is decreased, the Wenzel state is destabilized.
    (b) This destabilization of the Wenzel state is manifested not only in an increase in the slope of region I, but also in a corresponding increase in the slopes of the non-classical regions II - IV, from negative towards zero to eventually being positive.
    As a result, a decrease in pressure shifts the location of the barrier (black circles) to higher $\rhon$ and leads to a concomitant decrease in the height of the barrier.
    (c) The barriers for the wetting and dewetting transitions are shown here as a function of $\Delta P$ (simulation=solid lines, theory=dashed lines). 
    Both the simulated and the classical Cassie-to-Wenzel barriers (blue) decrease on increasing pressure, eventually disappearing at the intrusion pressure, $\Delta P_{\rm int}$.
    On the other hand, while the classical Wenzel-to-Cassie barrier (magenta) is independent of pressure, simulations suggest that the barrier to dewetting disappears at a sufficiently small extrusion pressure, $\Delta P_{\rm ext}$.
    (d) Pressure dependent hysteresis curves for $\rhon$, assuming the system remains in its metastable basin and is unable to surmount barriers larger than 1~$\kT$. 
     }
\end{figure}

%%%%%%%%%%%%%%%%%%%%%%%%%%%%%%%%%%%%
%%%
The implications of this non-classical pathway on the pressure dependence of the dewetting transition are even more interesting.
Because pressure favors configurations with higher densities in a well-defined manner, its effect on $\Delta F(\rhon)$ can be readily estimated, as shown in the Supporting Information.
As shown in Figure~3a, the most striking effect of changing pressure is seen in the slope of region I; as pressure is increased, the slope decreases and the Cassie state is destabilized.
The increase in pressure also leads to a decrease in the barrier to transition from the Cassie to the Wenzel state, as shown in Figure~3c.
The pressure at which this barrier for wetting disappears, $\Delta P_{\rm int}$, corresponds to the limit of stability (or spinodal) of the Cassie state; at $\Delta P = \Delta P_{\rm int}$, a system in the Cassie state will spontaneously descend into the Wenzel state, as shown in the Cassie-to-Wenzel hysteresis curve (blue) in Figure~3d.
The exact agreement between the theoretical and simulated $\Delta P_{\rm int}$-values seen in Figure~\ref{fig:pressure}c,d is not coincidental, but a consequence of the fact that the value of vapor-liquid surface tension, $\gamma_{\rm fit}$, used in the macroscopic theory, was obtained by fitting the simulated free energy profile in region I; see Supporting Information.
The reasonable agreement between $\gamma_{\rm fit}$ and the reported value of $\gamma_{\infty}$ for SPC/E water nevertheless suggests that the macroscopic theory prediction of intrusion pressure, $\Delta P_{\rm int} = -\gamma\cos\theta A_{\rm side}/V$~\cite{Patankar:2010df,butt2014super}, should be reasonably accurate even for surfaces with nanoscale texture, as suggested by recent experiments~\cite{Checco:PRL:2014}.
This success of macroscopic theory in describing $\Delta P_{\rm int}$, the pressure at which superhydrophobicity fails, is a direct consequence of its ability to capture region I of $\Delta F(\rhon)$ accurately.
%
%%%

%%%
While regions II to IV play no role in determining $\Delta P_{\rm int}$, their role is profoundly important in the reverse process, that is, the Wenzel to Cassie (dewetting) transition.
As pressure is decreased, not only does the slope of region I increase (Figure~3a), but the slopes of regions II - IV that are negative at $\Delta P=0$, also increase, approaching zero at sufficiently negative pressures.
Figure~3b zooms in on the liquid basin of $\Delta F(\rhon)$, and highlights that as pressure is decreased, the location of the peak in the free energy shifts
to higher $\rhon$, and is accompanied by a gradual decrease in the height of the dewetting barrier (Figure~3c).
This decrease in the dewetting barrier with decreasing pressure is in stark contrast with the macroscopic expectation that a constant adhesion barrier must be overcome to go from the Wenzel to the Cassie state~\cite{Patankar:Langmuir:2004}.
Under sufficient tension (negative pressure), the barrier goes to zero as the Wenzel basin reaches the limit of its stability.
Our simulations thus suggest that superhydrophobicity can be recovered, that is, a system in the Wenzel state can spontaneously and remarkably transition back into the Cassie state below a so-called ``extrusion'' pressure, $\Delta P_{\rm ext}$ (Figure~3d).

%---------------------------------figure3:Pressure-------------------------------------------------
\vspace{-0.3in}
%--------------------------------------------------------------------------------------------------------------------
%
%

 %---------------------------------figure4: Novel Surface-------------------------------------------------
\begin{figure*}[t]
    \begin{center}
    \includegraphics[width=0.99\textwidth]{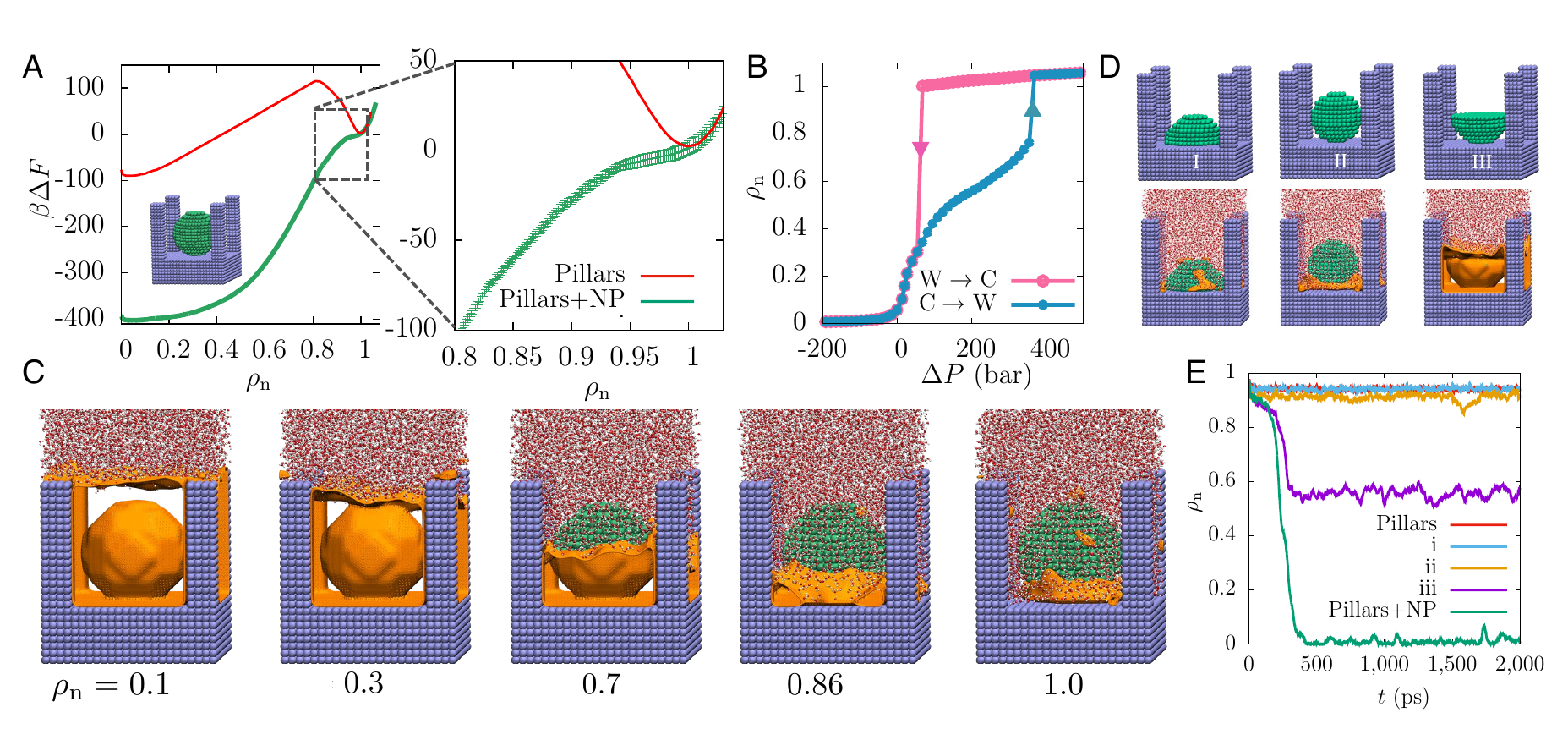}
    \end{center}
    \vspace{-0.15in}
    \caption{ 
     \label{fig:novel}
    Designing surface textures for spontaneous recovery of superhydrophobicity.
    (a) The free energetics of a surface with a 3.5~nm diameter spherical nanoparticle at the center of the cell (inset) are compared to the corresponding pillared surface; the surface modification stabilizes the superhydrophobic Cassie state and renders the Wenzel state unstable.
 (b) The modified surface features a positive extrusion pressure, $\Delta P_{\rm ext}=68$~bar, suggesting spontaneous recovery of superhydrophobicity at ambient conditions.
 (c) Representative instantaneous interface configurations highlight that dewetting commences at the center of the cell and spreads outwards, facilitating the spontaneous formation of the vapor-liquid interface, which then rises along the pillars.
    While waters stick to the top surface of the nanoparticle, the unfavorable gradient in water density between the edges and the center of the cell facilitates the depinning of the vapor-liquid interface.
(d) We simulate three additional variants of the pillared surface, which displace waters from the center of the cell using a: (i) a 3.5 nm hemispherical particle, (ii) a 3 nm spherical nanoparticle, and (iii) an inverted 3.5 nm hemispherical particle.
The three surfaces (top row) each violate one of the three criteria for spontaneous dewetting outlined in the text.
When initialized in the Wenzel state, these surfaces are unable to undergo complete dewetting, as evidenced by the final configurations of 2~ns long unbiased simulations (bottom row).
(e) The time dependence of $\rhon$ for such unbiased simulations highlights that the Wenzel state is unstable only for the pillared system with a 3.5 nm spherical nanoparticle.
\vspace{-0.2in}
}
  \end{figure*}
%--------------------------------------------------------------------------------------------------------------------

\vspace{0.2in}
%%%%%%%%%%%%%%%%%%%%%%%%%%%%%%%%%%%%
\section{Informing the Design of Robust Superhydrophobic Surfaces} 
%%%%%%%%%%%%%%%%%%%%%%%%%%%%%%%%%%%%
%%%
For the surface studied here, the extrusion pressure, $\Delta P_{\rm ext} = -300$~bar, is quite small.
While water can sustain significant tension, and experiments have been able to access $\Delta P\lesssim-1200$~bar before cavitation occurs~\cite{Caupin:PNAS:2014}, from a practical standpoint, superhydrophobic surfaces with significantly larger $\Delta P_{\rm ext}$ values that exceed atmospheric pressure are desirable.
Clues for designing such surfaces are contained within Figure~2, which suggests that it is hardest to remove water molecules from the center of the simulation cell. 
To destabilize those waters and facilitate the formation of the vapor-liquid interface, we modify the pillared morphology by adding a spherical nanoparticle at the center of the cell; see inset of Figure~4a.
A comparison of the free energy profile, $\Delta F(\rho_{\rm n})$, of this novel surface with that of the pillared surface (Figure~4a) highlights that the introduction of the nanoparticle has dramatic influence on $\Delta F(\rhon)$; the Wenzel state is destabilized to such an extent that it is no longer metastable, but has been rendered unstable at $\Delta P=0$.
%%%
%%%
In fact, as shown by the hysteresis curves in Figure~4b, the Wenzel state remains unstable up to an extrusion pressure, $\Delta P_{\rm ext}=+68$~bar.
The pressure dependence of the free energy profiles and the barriers to wetting/dewetting are included in the Supporting Information.
Thus, a system prepared in the wet (Wenzel) state, by condensation or by starting at a high pressure for example, should spontaneously dewet on decreasing the pressure below 68 bar, and lead to a recovery of the superhydrophobic Cassie state.
To investigate this possibility and examine whether barriers exist in co-ordinates orthogonal to $\rhon$, we performed 26 equilibrium simulations starting with $\rhon \approx1$.
In each case, the system spontaneously descends into the Cassie state within 700~ps, indicative of a barrierless transition from the Wenzel to the Cassie state.
Details of these simulations are provided in the Supporting Information and one of the dewetting trajectories is included as a Supporting Movie.
%%%

%%%
The dewetting trajectory as well as characteristic configurations along the dewetting pathway (Figure~4c), highlight that the strongly confined region between the nanoparticle and the basal surface nucleates a vapor bubble ($\rhon \approx1$), which then grows to facilitate the spontaneous formation of an intact vapor-liquid interface ($\rhon=0.83$).
Once formed, the vapor-liquid interface begins to rise, and although it sticks at the top of the nanoparticle, the interface continues to rise along the pillars; the unfavorable intra-cell density gradient then facilitates the barrierless depinning of the interface.
This dewetting pathway suggests that the spontaneous recovery of superhydrophobicity requires:
(i) a confined (negative curvature) region of nanoscopic dimensions to nucleate a vapor bubble, which 
(ii) must grow to a large enough size to facilitate the formation of an intact vapor-liquid interface, which in turn
(iii) must not be pinned by surface features as it rises along the pillars.
As shown in the Supporting Information, by following these general principles, we were able to propose a second surface texture design with an unstable Wenzel state.
We also study three additional surfaces that respectively violate one of the above criteria (Figure~4d).
By running unbiased simulations initialized in the Wenzel state, we show that none of the three surfaces are able to spontaneously transition to the Cassie state (Figure~4e).
The fact that each of the above criteria must be satisfied for spontaneous dewetting, could explain why nanotextured surfaces that spontaneously recover superhydrophobicity have not been discovered serendipitously, despite the use of both nanoparticles and nanoscale pillars to texture surfaces~\cite{xu2012transparent,Checco:PRL:2014}. 
%
%%%

\vspace{-0.1in}
%%%%%%%%%%%%%%%%%%%%%%%%%%%%%%%%%%%%
\section{Outlook and Discussion} 
%%%%%%%%%%%%%%%%%%%%%%%%%%%%%%%%%%%%
%%%
%
The last decade has seen an explosion of studies pertaining to superhydrophobic surfaces, ranging from the development of novel techniques for their synthesis, to those aimed at understanding the properties of these surfaces and their putative applications.
Our work informs both fundamental and applied aspects of superhydrophobicity, not only uncovering non-classical Cassie--Wenzel transition pathways, but also suggesting strategies for the rational design of robust superhydrophobic surfaces, which can spontaneously recover their superhydrophobicity.
In particular, we show that the free energetics of wetting-dewetting transitions on nanotextured surfaces, the corresponding mechanistic pathways, and their dependence on pressure, are all strongly influenced by collective water density fluctuations.
While the importance of fluctuations at the nanoscale is clear, the extent to which fluctuations influence dewetting pathways at larger texture sizes, remains an interesting open question.
We note, however, that non-classical dewetting effects stem from the stabilization of the dewetted morphologies, which precede the formation of an intact vapor-liquid interface at the basal surface; once a vapor-liquid interface is nucleated, the remainder of the dewetting process is classical.
Because the width of this nascent vapor-liquid interface is not expected to depend on the texture size, and should always have nanoscale dimensions, fluctuations may continue to play an important role in dewetting surface textures with significantly larger feature sizes.
Our results also clarify that while certain aspects of dewetting on nanotextured surfaces are non-classical, other aspects can be classical.
In particular, because the intrusion pressure depends on the free energetics of the (classical) ascent of the vapor-liquid interface along the pillars, it is well-described by macroscopic theory.
Similarly, macroscopic models~\cite{cassie,wenzel} have also been shown to capture apparent contact angles of liquid droplets localized in the Cassie or Wenzel basins~\cite{Kumar:JCP:2011, Leroy:Langmuir:2012, Shahraz:Langmuir:2013, Kumar:Langmuir:2013}.
An understanding of the Cassie--Wenzel transition pathways also facilitates the rational design of surfaces with superior superhydrophobicity.
While numerous methods have been suggested for texturing hydrophobic surfaces, from top-down fabrication to bottom-up self-assembly techniques~\cite{synthesis-review,Tuteja:2007fh,kang2014rapid}, a key bottleneck in their widespread adoption has been the fragility of the superhydrophobic Cassie state and the associated irreversible loss of superhydrophobicity upon wetting~\cite{Bormashenko-review}.
Our findings represent a major step in addressing this challenge, highlighting the importance of water density fluctuations in stabilizing non-classical pathways on nanotextured surfaces, which reduce dewetting barriers and enable the spontaneous recovery of superhydrophobicity.
Our results also suggest a general strategy for augmenting the design of existing nanotextured surfaces.
By identifying the pathways to dewetting, and in particular, the regions that are hardest to dewet, we were able to inform the location of sites, where the introduction of additional texture would be optimal. 
In conjunction with recent advances in introducing texture at the nanoscale~\cite{rahmawan2013self,Checco:PRL:2014}, our results thus promise to pave the way for robust superhydrophobic surfaces with widespread applicability under the most challenging conditions.
One example of such an application involves sustained underwater operation~\cite{cottin2003low}, which would require the surface texture to not only remain dry under hydrostatic pressure, but also be able to return to the dry state, if the texture wets in response to a perturbation.
Another example pertains to condensation heat transfer~\cite{azimi2013hydrophobicity}, wherein an unstable Wenzel state would facilitate the immediate roll-off of condensing water droplets, and enable drop-wise condensation to be sustained at higher fluxes.
\vspace{-0.1in}
\section{Materials and Methods} 
\subsection{Simulation Setup}
Each of the surfaces that we study is composed of atoms arranged on a cubic lattice with a lattice spacing of 0.25~nm; the surface atoms are constrained to remain in their initial positions throughout the simulations.
The basal surface, which is situated at the bottom of the simulation box, is made of 8 layers of atoms and is 2~nm thick.
On the pillared surface, pillars of height, $H$=4.5~nm, and a square cross-section with width, $W$=2~nm, are employed to introduce surface texture.
The pillars are placed on a square lattice with inter-pillar spacing, $S$=4~nm. 
We also study surfaces that additionally contain spherical or hemispherical nanoparticles. 
The nanoparticles are also made of atoms on a cubic lattice with a lattice spacing of 0.25~nm, and are placed at the center of the simulation cell, touching the top of the basal surface.
Each system is periodic in all three spatial directions and is initialized in the Wenzel state.
The surfaces are hydrated with roughly 7,000 water molecules, so that even in the Wenzel state, a 2~nm thick water slab rests above the pillars.
We also provide a buffering vapor layer at the top of the simulation box, which is roughly 6~nm thick in the Wenzel state.
As waters leave the textured region, the buffering vapor layer shrinks to roughly 2~nm in the Cassie state.
The buffering vapor layer, which is thus present in our system at all times, ensures that water is always in coexistence with its vapor, and correspondingly, that the system pressure is equal to the water-vapor coexistence pressure at 300~K~\cite{Miller:PNAS:2007}. 
To break translational symmetry, and ensure that the vapor layer remains at the top of the 6$\times$6$\times$15~nm$^3$ simulation box, we include a repulsive wall at $z$=14.5~nm.

\subsection{Simulation Details}
We have chosen the SPC/E model of water~\cite{SPCE} because it adequately captures the experimentally known features of water, such as surface tension, isothermal compressibility, and the vapor-liquid equation of state near ambient conditions, all of which are important in the study of dewetting on hydrophobic surfaces~\cite{Chandler:Nature:2005,vega:2007,varilly2013water,LCW}.
The surface atoms interact with the water oxygens through the Lennard-Jones (LJ) potential ($\sigma$=0.35~nm, $\epsilon$=0.40~kJ/mol).
As shown in the Supporting Information, this choice leads to a flat surface water droplet contact angle, $\theta$=116.2$^\circ$, in accord with contact angles observed on typical hydrophobic surfaces, such as alkyl-terminated self-assembled monolayer surfaces~\cite{Godawat:PNAS:2009}.
We use the GROMACS molecular dynamics (MD) simulation package~\cite{gmx4ref}, suitably modified to perform INDUS simulations in the canonical (NVT) ensemble. 
A detailed description of the INDUS calculations~\cite{Patel:JPCB:2010,Patel:JSP:2011}, which we use to characterize the free energetics of wetting-dewetting transitions on nanotextured surfaces, is included in the Supporting Information.
To maintain a constant temperature, $T$=300~K~\cite{Bussi:JCP:2007}, the canonical velocity-rescaling thermostat with a time constant of 0.5~ps is employed.
LJ interactions and the short-ranged part of the electrostatic interactions are truncated at 1~nm, and the Particle Mesh Ewald algorithm is employed to treat the long-ranged part of electrostatic interactions~\cite{PME}. 
The SHAKE algorithm is used to constrain the bond lengths of the water molecules~\cite{SHAKE}.
\vspace{-0.1in}
%\end{materials}
%
\begin{acknowledgments}
A.J.P. gratefully acknowledges financial support from the National Science Foundation (UPENN MRSEC DMR 11-20901 and CBET 1511437), and would like to thank John Crocker for helpful discussions.
\end{acknowledgments}
%

%\end{article}

%\bibliography{MakeArxivMerged}

\clearpage
\newpage

\title{Supporting Information for ``Spontaneous Recovery of Superhydrophobicity on Nanotextured Surfaces"}

%Author List
\author{Suruchi Prakash}
\affiliation{Department of Chemical \& Biomolecular Engineering, University of Pennsylvania, Philadelphia, PA 19104, USA}

\author{Erte Xi}
\affiliation{Department of Chemical \& Biomolecular Engineering, University of Pennsylvania, Philadelphia, PA 19104, USA}

\author{Amish J. Patel}
\email{amish.patel@seas.upenn.edu}
\affiliation{Department of Chemical \& Biomolecular Engineering, University of Pennsylvania, Philadelphia, PA 19104, USA}
%Date
\date{\today}

\maketitle
\raggedbottom
%
%\begin{article}
%****************************************** New Section *********************************************************%
\bibliographystyle{pnas}
\section{Supporting Methods: Indirect Umbrella Sampling (INDUS) Calculations}
The INDUS method entails biasing the number of waters, $N$, in a volume, $V$, of interest, indirectly through a coarse-grained water number, $\Ntild$, that is closely related to $N$.
Because $N$ changes discontinuously when a particle crosses the boundary of $V$, biasing $N$ can lead to impulsive forces.
In contrast, $\Ntild$ is a continuous function of the particle positions. 
To coarse-grain $N$, we use a Gaussian function with a standard deviation of $0.01~\nm$ that is truncated at $0.02~\nm$ and shifted down; this choice of parameters allows the resultant $\Ntild$ and $N$ to remain strongly correlated.
To sample $N$ over the entire range between the Cassie and the Wenzel basins, we run a series of simulations with different harmonic biasing potentials $U_j(\Ntild) = \frac{\kappa}{2} (\Ntild - \Ntild_j^*)^2~(j=1,2,...,n_{\rm w})$, with $\kappa=0.03$~kJ/mol and the $\Ntild_j^*$-values chosen to allow sufficient overlap between adjacent windows.
The INDUS simulations enable us to sample the biased joint distribution function, $P_V^j(N,\Ntild)$, which represents the probability of observing $N$ waters and $\Ntild$ coarse-grained waters in $V$, in the presence of the biasing potential, $U_j(\Ntild)$.
We then unbias and stitch together the $n_{\rm w}$ biased joint distribution functions using the weighted histogram analysis method (WHAM)~\cite{roux_wham,Patel:JSP:2011}. 
Finally, we integrate the resultant unbiased joint distribution function over $\Ntild$ to obtain the probability, $P_V(N)$, of observing $N$ waters in $V$.
To compare partially wet states across different surfaces, we define a normalized water density, $\rhon \equiv N/N_{\rm liq}$, with $N$ being the number of water molecules in $V$ in the partially wet state, and $N_{\rm liq}$ being the corresponding value in the fully wet (Wenzel) state; that is, $N_{\rm liq}$ corresponds to the location of the liquid basin maximum in $P_V(N)$.
The free energy, $\Delta F(\rho_{\mathrm{n}})$, of the system in a partially wet state with normalized density $\rho_{\rm n}$, relative to that in the Wenzel state, which we report in the main text, is then obtained as:
\begin{equation}
    \beta\Delta F(\rho_{\mathrm{n}})  = - \ln \big[ P_V(N) / P_V(N_{\mathrm{liq}}) \big],
    \label{eq:free}
\end{equation}
where $\beta^{-1}=\kT$ is the thermal energy, and $k_{\rm B}$ is the Boltzmann constant. 
%
%$$$$$$$$$$$$$$$$$$$$$$$$$$$$$$$$$New Section $$$$$$$$$$$$$$$$$$$$$$$$$$$$$%
\section{Supporting Discussion}
\subsection{Macroscopic Theory}
\begin{figure*}[tb]
   \centering
    \includegraphics[width=0.45\textwidth]{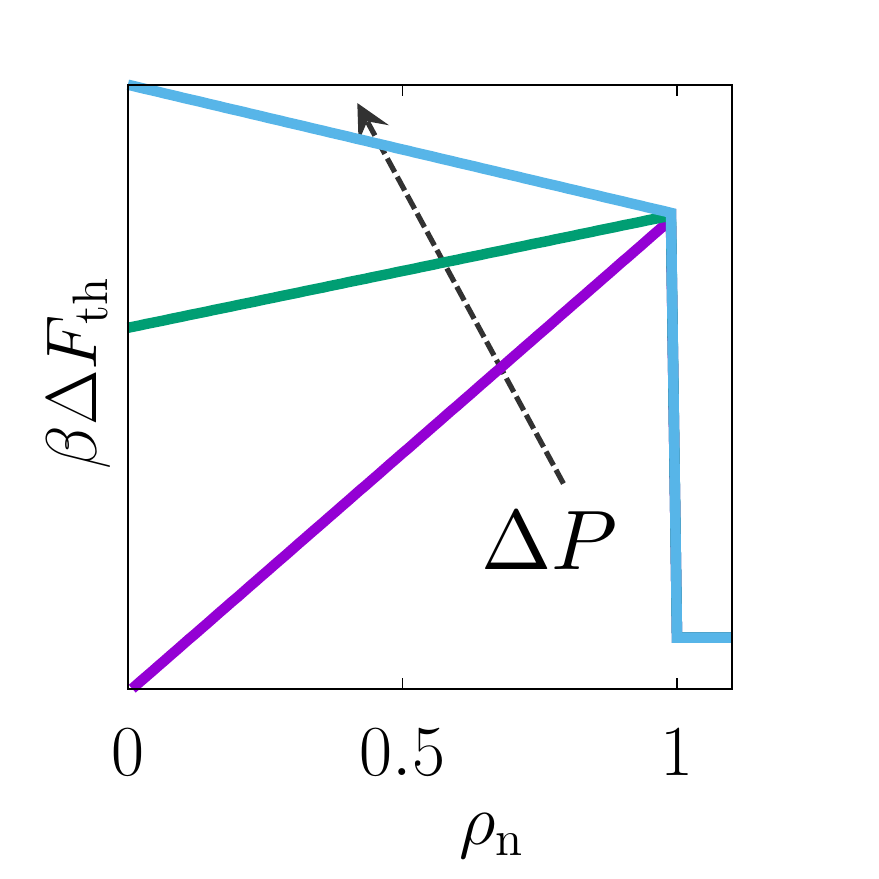}
    \caption{  
    Macroscopic theory predicts a barrier to the dewetting Wenzel-to-Cassie transition; the barrier at $\rhon=1$ corresponds to the work of adhesion, $\Delta F_{\rm adh}$, that must be done to nucleate a vapor-liquid interface at the basal surface. 
    The adhesion barrier is expected to be independent of pressure; $\Delta P$ only affects the slope of the linear region between the barrier and the Cassie state. 
    As pressure is increased, the slope of this region decreases, stabilizing the Wenzel state relative to the partially wet states. 
    For $\Delta P>\Delta P_{\rm int}$,  the Cassie state becomes unstable (blue curve).
     }
    \label{fig:macro}
\end{figure*}

According to macroscopic interfacial thermodynamics~\cite{Pat0408}, the free energy of a system in a partially wet state with normalized density $\rho_{\rm n}$, relative to that in the Wenzel state is:
\begin{align}
\Delta F_{\rm th}(\rhon) = \gamma A_{\rm vl}(\rhon)+\Delta \gamma A_{\rm sv}(\rhon) + \Delta PV( 1 - \rhon),
    \label{eq:macro1}
\end{align}
where $\gamma$ is the vapor-liquid surface tension, and $\Delta\gamma$ is the difference between the solid-vapor and solid-liquid surface tensions; according to the Young's equation, $\Delta\gamma=\gamma\cos\theta$. 
$A_{\rm vl}$ and $A_{\rm sv}$ are the vapor-liquid and the solid-vapor interfacial areas, respectively, in the partially wet state, and $\Delta P$ is the difference between the system pressure and the co-existence pressure at the system temperature, $T$.
%%%
Assuming that a partially wet state on the pillared surface is characterized by a flat vapor-liquid interface at a height $h$, above the base of the pillars, it follows that $h(\rhon) = H (1- \rhon)$~\cite{Pat0408,NosBhu0802}.
Then, $A_{\rm vl} = S ( S + 2W ) \equiv A_{\rm base}$,  is independent of $h$, whereas $A_{\rm sv}=A_{\rm base} + A_{\rm side} ( h / H )$, increases linearly with $h$. Here, $A_{\rm side}=4WH$ is the area of the vertical faces of the pillars. 
$\Delta F_{\rm th}(\rhon)$ is then given by:
\begin{equation}
\Delta F_{\rm th}(\rhon) = \Delta F_{\rm adh} + [ \gamma\cos\theta A_{\rm side} + \Delta P V ]\frac{h(\rhon)}{H},
    \label{eq:macro2}
\end{equation} 
where $\Delta F_{\rm adh} \equiv \gamma A_{\rm base} (1+\cos\theta)$ is always unfavorable (positive), and corresponds to the work of adhesion for creating the vapor-liquid interface.
Because $\cos\theta < 0$ for hydrophobic surfaces, the second term could be favorable (negative) if $\Delta P$ is sufficiently small, that is, if $\Delta P \le \Delta P_{\rm int} \equiv -\gamma\cos\theta A_{\rm side}/V$.
The two main features in $\Delta F_{\rm th}(\rhon)$ are clearly seen in Figure~\ref{fig:macro}; the first is a large adhesion barrier, $\Delta F_{\rm adh} $, at $\rhon=1$ that must be overcome to nucleate the vapor-liquid interface, and the second is the linear portion of $\Delta F_{\rm th}(\rhon)$ corresponding to the vapor-liquid interface rising along the pillars as $\rhon$ decreases.
As the pressure is increased, the slope of the linear portion decreases and partially wet states become less favorable compared to the Wenzel state.
At sufficiently large pressure ($\Delta P>\Delta P_{\rm int}$), the Cassie state can become unstable. 
%
%$$$$$$$$$$$$$$$$$$$$$$$$$$$$$$$$$New Section $$$$$$$$$$$$$$$$$$$$$$$$$$$$$%
%
%
\subsection{Comparison between Molecular Simulations and Macroscopic Theory: Uncovering the Importance of Fluctuations}
The partially wet configurations observed in our simulations for $0.2<\rhon<0.8$ contain a vapor-liquid interface that is perpendicular to the pillars, and at average heights that are consistent with the macroscopic theory approximation, $h(\rhon) = H (1- \rhon)$; see Figure~\ref{fig:havg}.
The dependence of the free energy profile on density is also classical for $0.2<\rhon<0.8$; that is, $\Delta F(\rhon)$ varies linearly with $\rhon$.
According to Equation~\ref{eq:macro2}, the slope of this linear region for $\Delta P=0$ is expected to be $-\gamma\cos\theta A_{\rm side}$.
Fitting the simulation data in that region allows us to extract an estimate of the vapor-liquid surface tension, $\gamma_{\rm fit} =67$~mJ/m$^2$, which is in reasonable agreement with the reported value of the vapor-liquid surface tension of SPC/E water~\cite{vega:2007}, $\gamma_{\rm SPC/E} =63.6$~mJ/m$^2$. 
In order to obtain $\gamma_{\rm fit}$ from the fitted slope, we use $\cos \theta=-0.44$ and effective values of $W=2.18$~nm and $H=4.62$~nm, for reasons explained in Figures~\ref{fig:ca} and~\ref{fig:offsets}, respectively.
Note that this value of $\gamma_{\rm fit}$ is used in all comparisons to macroscopic theory in the main text. 
In particular, using the value of $\gamma_{\rm fit}$, macroscopic theory can be used to predict the dewetting barrier, $\Delta F_{\rm adh} =\gamma A_{\rm base} (1+\cos\theta)$, and to obtain the entire free energy profile, as shown in Figure~\ref{fig:free} (black solid line).
It is clear that macroscopic theory significantly overpredicts the barrier for dewetting ($\approx 270 ~\kT$) as compared to the simulations ($\approx 118 ~\kT$).
We now describe how we obtain estimates of the contact angle, $\theta$, and refined estimates of the geometric parameters, $W$, $H$ and $S$, in order to facilitate a comparison of the free energy profile, $\Delta F(\rhon)$, obtained from our simulations with the corresponding prediction of macroscopic theory, $\Delta F_{\rm th}(\rhon)$ (Equation 3).
%
%
%****************************************** CONTACT ANGLE*********************************************************%
\subsubsection{Contact Angle}
To estimate the water droplet contact angle on the flat basal surface, we simulate a box-spanning $12$~nm long cylindrical water droplet on a $18~\nm\times12~\nm\times2~\nm$ surface; a simulation snapshot is shown in Figure~\ref{fig:ca}a.
Starting with a cuboid shaped droplet (6~nm$\times12$~nm$\times6$~nm; 14,353 water molecules) on the surface, we first equilibrate the system for 3~ns. 
We then run the simulation for another 5~ns, and use the positions of the water oxygens every 1~ps to determine an averaged two-dimensional map of the water density; such a map is shown in Figure~\ref{fig:ca}b.
The $z$-coordinate in the averaged water density map corresponds to the distance of water oxygens from the first layer of surface atoms and the $x$-coordinate represents their distance from the $x$-component of the center of mass of the droplet.
To account for any axial undulations in the shape of the 12~nm long cylindrical droplet, we divide the droplet into twelve 1~nm thick cylindrical slabs, and correct the $x$-coordinate of the water oxygens with respect to the center of mass of the corresponding slab.
Once the averaged water density map is computed, we determine its half bulk-density isosurface, as shown in Figure~\ref{fig:ca}c.
To estimate the contact angle, we then fit the isosurface to a circle of radius, $R$, with its center at a height, $z_{\rm cen}$,  above the surface, using the equation:
\begin{equation}
    x = \sqrt{R^2-(z-z_{\rm cen})^2}
    \label{eq:ca1}
\end{equation}
The fitted circle is also shown in Figure~\ref{fig:ca}c (black line) and is in excellent agreement with the data.
The contact angle, $\theta$, is obtained as the angle between the tangent to the fitted circle at $z=0.7$~nm and the surface, and is given by: 
\begin{equation}
    \cos\theta  = -(z_{\rm cen}-0.7)/R
    \label{eq:ca2}
\end{equation}
The fit parameters, $z_{\rm cen}=2.34$~nm and $R=3.71$~nm,  yield a contact angle, $\theta=116.2^\circ \pm 0.2^\circ$. 
The standard error in the contact angle, reported above, is obtained from independent estimates of $\theta$ for each of the twelve 1~nm thick cylindrical slabs described above.
Note that the particular value of the contact angle that we obtain depends on the (somewhat arbitrary) choice of 0.7~nm for the height of 3-phase contact line; however, an uncertainty of $0.1$~nm in the height of the contact line, translates to an additional uncertainty of only $1.7^\circ$ in the contact angle.
%
%****************************************** Offsets*********************************************************%
\subsubsection{Offsets in Geometric Parameters}
The values of the pillar width $W$, its height $H$ and the inter-pillar separation $S$, reported in the main text, correspond to the distances between the centers of the corresponding surface atoms. 
To compare our simulations with macroscopic theory, here we obtain more refined estimates of these geometric parameters.
In particular, we use the fact that the effective volume of the textured region, $V_{\rm eff}$, ought to be related to $N_{\rm liq}$ (the number of waters in $V$ with the system in the liquid basin), through the simple expression, $V_{\rm eff}= N_{\rm liq}/\rho_{\rm B}$, where $\rho_{\rm B}$ is the bulk liquid density.
We estimate $N_{\rm liq}$ (and thereby, $V_{\rm eff}$) by running equilibrium simulations for 10 different geometries ($S = 3, 4$~nm and $H = 2, 3, 4, 4.5, 5$~nm).
We then assume that $V_{\rm eff}$ is related to effective values of the geometric parameters as $V_{\rm eff} = S_{\rm eff}(S_{\rm eff}+2W_{\rm eff}) H_{\rm eff}$.
Consistent with this assumption, we find that $N_{\rm liq}$ is a linear function of $H$ for both $S=3 ~\nm$ and $S= 4~\nm$, as shown in Figure~\ref{fig:offsets}.
Further, the $x$-intercepts obtained from both the corresponding fits are identical, allowing us to estimate the effective height as $H_{\rm eff}=H+0.12$~nm.
The two fitted slopes are then used to obtain $S_{\rm eff} =S-0.21$~nm and $W_{\rm eff} =W+0.18$~nm respectively.
These effective values of $W$, $H$ and $S$ are used in all comparisons to macroscopic theory that are described in this section (Figure~\ref{fig:free}) as well as in the main text; the subscript `eff', however, is dropped from all equations for the sake of simplicity.
Note that the particular values of $\gamma_{\rm fit}$ and $\Delta F_{\rm adh}$ depend on the algorithm that we employ to estimate the offsets.
However, neither the algorithm employed for determining the offsets, nor the accounting of the offsets itself, alters our finding that there is a large discrepancy in the simulated and predicted values of the barrier separating the Cassie and the Wenzel states.
%
%********************************Interface calculations*****************************************************%
\subsection{Algorithm for Estimating Instantaneous Interfaces} 
Our procedure for calculating instantaneous interfaces closely follows that prescribed by Willard and Chandler~\cite{Willard:JPCB:2010}, and builds upon it.
In addition to using Gaussian smearing functions to obtain a spatially coarse-grained water density, we also incorporate a coarse-grained density of the surface atoms to obtain an overall normalized coarse-grained density as:
\begin{equation}
\tilde{\rho}_{\rm total}(x,y,z) = \frac{\tilde{\rho}_{\rm water}(x,y,z)}{\rho_{\rm B}} +  \frac{\tilde{\rho}_{\rm surface}(x,y,z)}{\rho^{\rm max}_{\rm surface}},
\label{eq:ii1}
\end{equation}
where $\tilde{\rho}_{\rm \alpha}$ is the coarse-grained density field of species $\alpha$, and $\rho^{\rm max}_{\rm surface}$ is the maximum in $\tilde{\rho}_{\rm surface}(x,y,z)$ for a flat surface consisting of 1 layer of surface atoms. 
$\tilde{\rho}_{\rm \alpha}$ is obtained using:
\begin{equation}
	\tilde{\rho}_{\alpha}(x,y,z) = \sum_{i=1}^{N_{\alpha}} \phi(x_i-x)\phi(y_i-y)\phi(z_i-z), 
\label{eq:ii2}
\end{equation}
where $\alpha$ represents either the water oxygen atoms or the surface atoms, $N_{\alpha}$ is the number of atoms of type $\alpha$, and $(x_i,y_i,z_i)$ correspond to the co-ordinates of atom $i$.
$\phi(x)$ is chosen to be Gaussian with standard deviation, $\sigma$, and is truncated at $|x|=c$, shifted down and normalized:
\begin{equation}
	\phi(x)=k[e^{-x^2/2\sigma^2}- e^{-c^2/2\sigma^2}] \Theta(c-|x|) ,
\label{eq:ii3}
\end{equation}
where $\Theta(x)$ is the Heaviside step function and $k$ is the normalization constant.
We choose $\sigma=0.16$~nm for the surface atoms, $\sigma=0.24$~nm for water oxygens, and $c=0.7$~nm for both.
We define the instantaneous interface encapsulating the dewetted regions as the $\tilde{\rho}_{\rm total}(x,y,z)=0.5$ isosurface, and employ the Marching Cubes algorithm to identify that isosurface~\cite{marching}.
While the precise location and the shape of the instantaneous interface depend on the values of the parameters employed to calculate it, we do not expect the qualitative insights obtained to be sensitive to the particular choice of parameters.
A detailed discussion of how the parameters affect the instantaneous interface calculation is beyond the scope of this work, and will be the subject of a separate publication.
\subsubsection{Instantaneous and Average Interfaces} 
Instantaneous interfaces encompassing the dewetted regions (orange) are shown in Figure~\ref{fig:ii}a for a range of $\rhon$-values spanning the Cassie and the Wenzel states. 
In our simulations, water molecules are present everywhere except in the dewetted regions; however, we do not show them here for the sake of clarity.
For the same $\rhon$-values, we also show the half-density isosurfaces of the time-averaged coarse-grained density field (Figure~\ref{fig:ii}b).
Both the average and the representative instantaneous interfaces highlight the dewetted morphologies corresponding to the different regions in the free energy profiles, namely: vapor-liquid interface perpendicular to the pillars for $\rhon<0.8$, a donut shaped dewetted region for $0.81<\rhon<0.83$, stripes of vapor for $0.83<\rhon<0.93$, and vapor pockets at the base of the pillars for $\rhon>0.93$. 
%
%********************************average height calculations*****************************************************%
\subsubsection{ Average Interface Heights}
For configurations with $\rhon <0.8$, the vapor-liquid interface is completely detached from the basal surface (Figure~\ref{fig:ii}).
To test the macroscopic theory approximation built into Equation~\ref{eq:macro2}, that is, $h(\rhon) = H(1-\rhon)$, here we obtain the average height of vapor-liquid interface as a function of $\rhon$.
For each configuration, we first obtain an average interface height by averaging over the $x$- and $y$- coordinates.
Before averaging over all configurations with a particular value of $\rhon$, care must be exercised in assigning appropriate weights to those configurations because they were obtained using biased simulations. 
To ensure this, we perform this average within the framework of WHAM~\cite{wham1,roux_wham}.
$\avg h$ thus obtained is plotted as a function of $\rhon$ in Figure~\ref{fig:havg} and agrees well with the macroscopic theory approximation.
%
%********************************Pressure*****************************************************%
\subsection{Pressure Dependence of $\Delta F(\rhon)$}
While our simulations are performed in the NVT ensemble, having a vapor-liquid interface in the simulation box ensures that the system is at its co-existence pressure~\cite{Miller:PNAS:2007}. 
The free energy profiles, $\Delta F(\rhon)$, that we estimate, are thus at $\Delta P = 0$.
However, increasing pressure favors configurations with higher densities in a well-defined manner; free energy profiles at other pressures can thus be readily estimated (to within an additive constant) by reweighting the simulation results obtained at the co-existence pressure.
The net result is the addition of a term linear in $\rhon$, analogous to the last term in Equation~\ref{eq:macro1}. 
Assuming the liquid basin to be the zero of energy at each pressure, we get:
$\Delta F(\rhon;\Delta P)=\Delta F(\rhon;\Delta P=0)+V\Delta P(1-\rhon)$.
\subsection{Spontaneously Recovering Superhydrophobicity} 
Here we describe the effect of pressure on the free energy profiles and on the magnitude of the wetting and dewetting barriers for the surface considered in the main text, which contains a 3.5 nm nanoparticle at the center of the pillared surface.
Using unbiased simulations initialized in the Wenzel state, we also explicitly demonstrate that this surface undergoes spontaneous dewetting at $\Delta P=0$. 
\subsubsection{Pressure Dependence of Free Energy Profiles and Barriers}
Figure~\ref{fig:pressureNP}a shows density-dependent free energy profiles at three different pressures.
At $\Delta P=0$, as discussed in the main text, the Wenzel state is not only metastable, but has been rendered unstable.
Higher pressures favors states with higher densities; as pressure is increased, the Wenzel state becomes metastable, and then stable above $\Delta P=255$~bar.
Figure~\ref{fig:pressureNP}b shows the barriers for the wetting (Cassie-to-Wenzel)  and dewetting (Wenzel-to-Cassie) transitions; the pressures where these barriers disappear correspond to the intrusion and extrusion pressures, respectively.
\subsubsection{Spontaneous Wenzel-to-Cassie Transition}
The positive extrusion pressure of this system implies that it should spontaneously dewet upon quenching from a fully wet state to ambient conditions ($\Delta P\approx0$).
To examine this possibility and investigate whether barriers exist in co-ordinates orthogonal to $\rhon$, we performed 26 unbiased simulations at $\Delta P=0$, starting from the fully wet state.
The initial positions of the waters were the same for each simulation, but their initial velocities were randomized.
In each case, the system spontaneously transitions to the Cassie state within 700~ps, suggesting a barrierless transition from the Wenzel to the Cassie state. 
$\rhon$ as a function of time is shown for one such trajectory in Figure~\ref{fig:timeNP}a.
A movie of the same trajectory, highlighting the evolution of the instantaneous interface encompassing the dewetted regions, has also been included as Supplementary Movie 1.
Finally, the distribution of times required to reach $\rhon =0.2$ across the 26 simulations is shown in Figure~\ref{fig:timeNP}b.
%
%********************************T-Sysyem**********************************************%
%
\subsection{Another Nanotextured Surface with Robust Superhydrophobicity}
In the main text, we showed that the transition state of the pillared system contains water molecules that stick to the center of the cell.
By introducing a large enough nanoparticle at the center of the cell, and destabilizing those water molecules, we were able to achieve spontaneous recovery of superhydrophobicity.
The dewetting pathway of this novel surface then revealed general guiding principles for making surfaces with an unstable Wenzel state.
In particular, we argued that: (i) a region of nanoscopic confinement is needed to nucleate a vapor bubble, (ii) the bubble must be able to grow spontaneously and become large enough to facilitate the formation of an intact vapor-liquid interface, and (iii) the interface must not be pinned by surface elements, as it rises along the pillars.
We further showed that variants of the robust surface that violate even one of these criteria, do not possess an unstable Wenzel state. 
We now show the pillared surface with a 3.5 nm nanoparticle is not unique; by introducing a `T' element (see Figure~\ref{fig:novel}, inset) with 
dimensions that were judiciously chosen to satisfy the criteria described above, we are able to similarly destabilize the Wenzel state.
The free energy profile, $\Delta F(\rho_{\rm n})$, of this surface is compared with that of the corresponding pillared surface in Figure~\ref{fig:novel}, and displays a Cassie state similar to the pillared system; however, the Wenzel state has once again been destabilized to such an extent that it is unstable at $\Delta P=0$.
We note that such surfaces have already been synthesized with microscale dimensions~\cite{Tuteja:2007fh}; however, they have not yet been synthesized at the nanoscale.
\section{Supporting Movie}
Spontaneous recovery of superhydrophobicity on the pillared surface with a 3.5~nm diameter spherical nanoparticle (Figure~\ref{fig:movie}). 
The first 380 ps of an unbiased simulation trajectory, starting from the wet Wenzel state, highlights the spontaneous recovery of the superhydrophobic Cassie state under ambient conditions ($\Delta P=0$).
Cavities (shown in orange) are nucleated in the highly confined region between the nanoparticle and the basal surface and grow outwards, facilitating the formation of the vapor-liquid interface, which then starts to rise along the pillars.
Although the interface is temporarily pinned at the top of the nanoparticle, it continues to rise along the pillars; the resulting unfavorable gradient in water density across the cell, leads to the barrierless depinning of the interface.
%
%\bibliography{superhydrophobic_SI}
%

%
%\end{article}
%\end{document}
%%%
%----------------------------------FIGURE 1------------------------------
%--------------------------------------------------------------------------------------------------------------------
%----------------------------------FIGURE 2: SIM-THEORY------------------------------------------------------
\begin{figure*}[tb]
   \centering
    \includegraphics[width=0.45\textwidth]{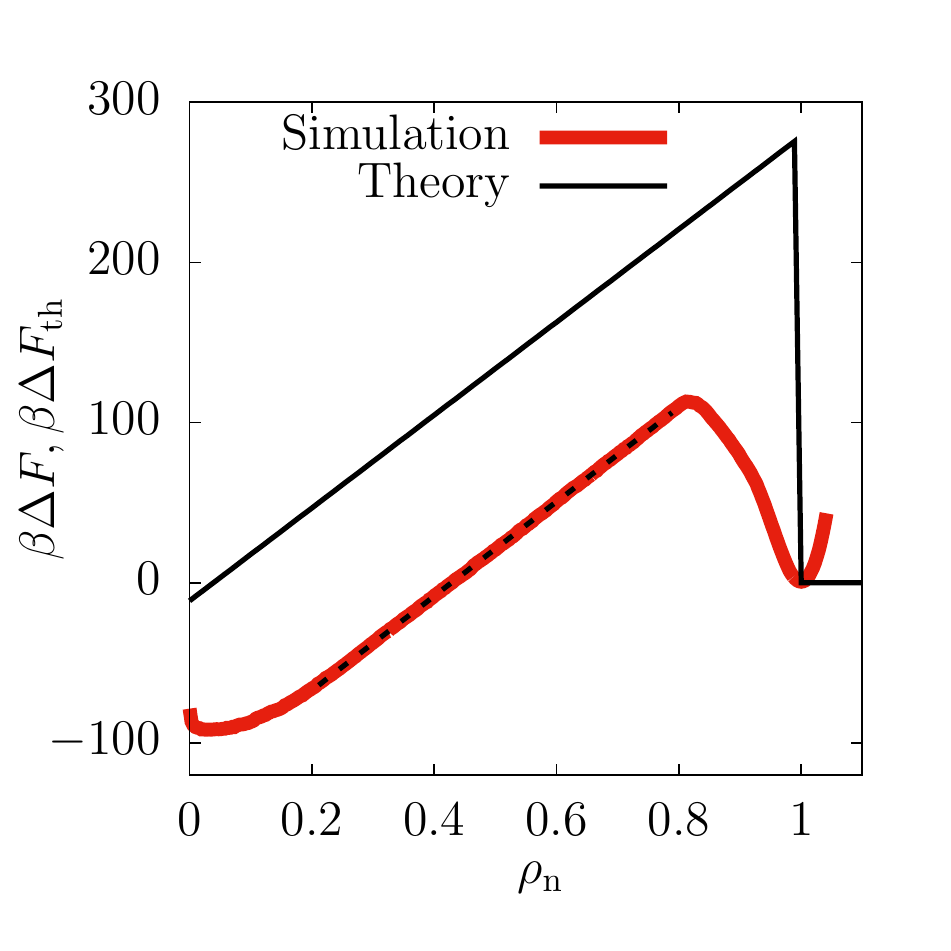}
    \caption{  
    Comparison of the simulated $\Delta F(\rhon)$ (red) with the corresponding macroscopic theory prediction, $\Delta F_{\rm th}(\rhon)$ (black) using $\gamma_{\rm fit}=67$~mJ/m$^2$.
    We extract $\gamma_{\rm fit}$ by fitting the simulation data in the linear region between $0.2<\rhon<0.8$ (dashed line), and comparing the fitted slope to $-\gamma\cos\theta A_{\rm side}$.    
    }
    \label{fig:free}
\end{figure*}
%--------------------------------------------------------------------------------------------------------------------
%----------------------------------FIGURE 3: Contact Angle------------------------------------------------------
\begin{figure*}[tb]
    \centering
    \includegraphics[width=1.0\textwidth]{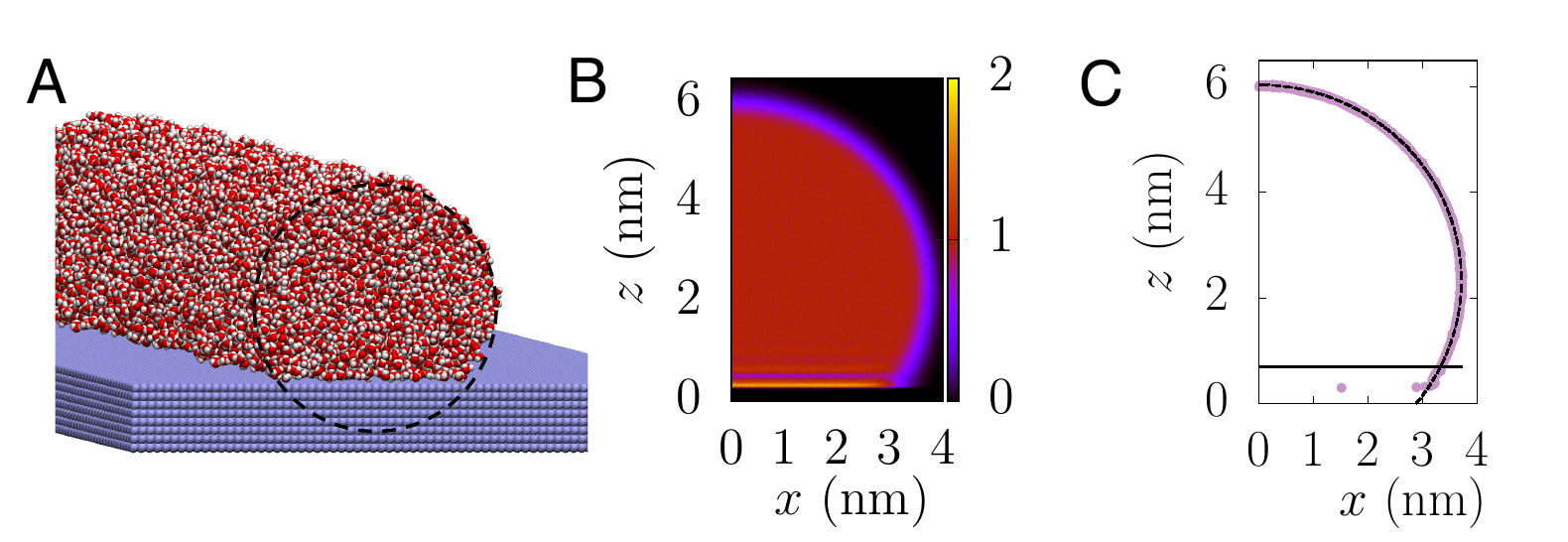}
    \caption{  
    To estimate the contact angle of the flat basal surface, we perform simulations of a cylindrical water droplet on the surface.
(a) Simulation snapshot of the cylindrical droplet. 
 (b) The average normalized density field as a function of distance from the $x$-component of the droplet center of mass and the height, $z$, of the droplet above the flat surface.
(c) The half-density isosurface is fit to a circle (Equation~\ref{eq:ca1}), and yields the parameters, $z_{\rm cen}=2.34$~nm and $R=3.71$~nm, which lead to $\theta=116.2^\circ$ (Equation~\ref{eq:ca2}).
    }
    \label{fig:ca}
\end{figure*}
%--------------------------------------------------------------------------------------------------------------------
%----------------------------------FIGURE 4 :Offsets------------------------------------------------------
\begin{figure*}[tb]
    \centering
    \includegraphics[width=0.45\textwidth]{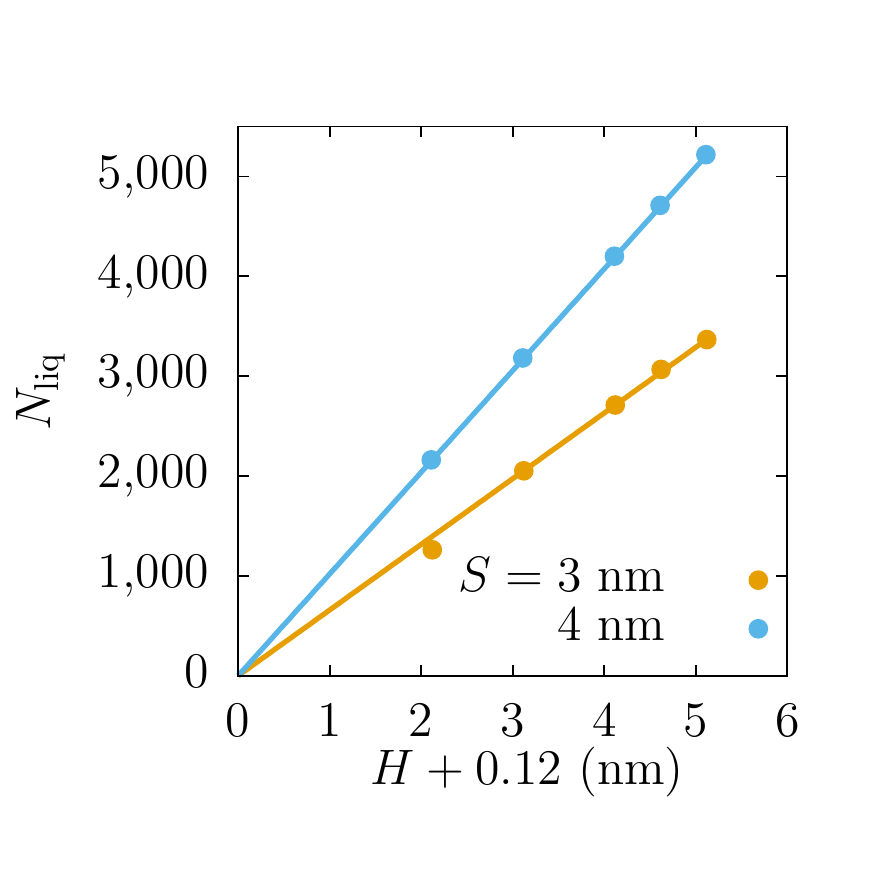}
    \caption{ 
    Linear fits to $N_{\rm liq}$ as a function of the pillar height, $H$, plus an offset of 0.12~nm, pass through the origin. 
    Thus, $H_{\rm eff}=H+0.12~\nm$.
    The slopes of the fitted lines enable us to similarly determine the offsets in pillar width, $W$, and inter-pillar separation, $S$.
    }
    \label{fig:offsets}
\end{figure*}
%--------------------------------------------------------------------------------------------------------------------
%----------------------------------FIGURE 5 :Interfaces------------------------------------------------------
\begin{figure*}[tb]
    \centering
    \includegraphics[width=1.0\textwidth]{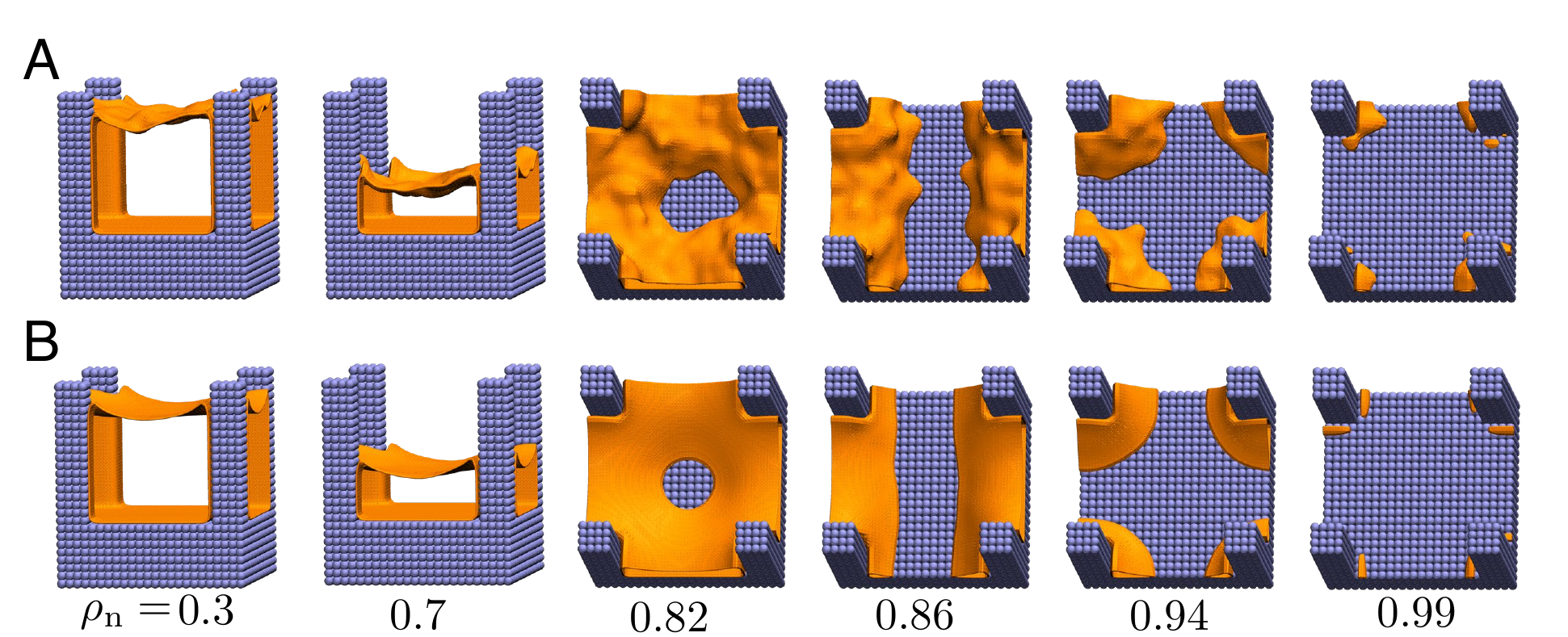}
    \caption{ 
    (a) Representative instantaneous interfaces and (b) average interfaces encompassing the dewetted regions are shown in orange, for partially wet states spanning the entire range of $\rhon$-values between the Cassie and Wenzel states.
Water molecules are not shown for the sake of clarity.
The configurations for $\rhon<0.8$ (side view) are classical with the vapor-liquid interface ascending along the pillars as $\rhon$ decreases.
In contrast, for $0.8<\rhon<1$ (top view), the system displays non-classical configurations with  at least three distinct dewetted morphologies.   
    }
    \label{fig:ii}
\end{figure*}
%--------------------------------------------------------------------------------------------------------------------
%
%----------------------------------FIGURE 6 :Average Height------------------------------------------------------
\begin{figure*}[tb]
    \centering
    \includegraphics[width=0.45\textwidth]{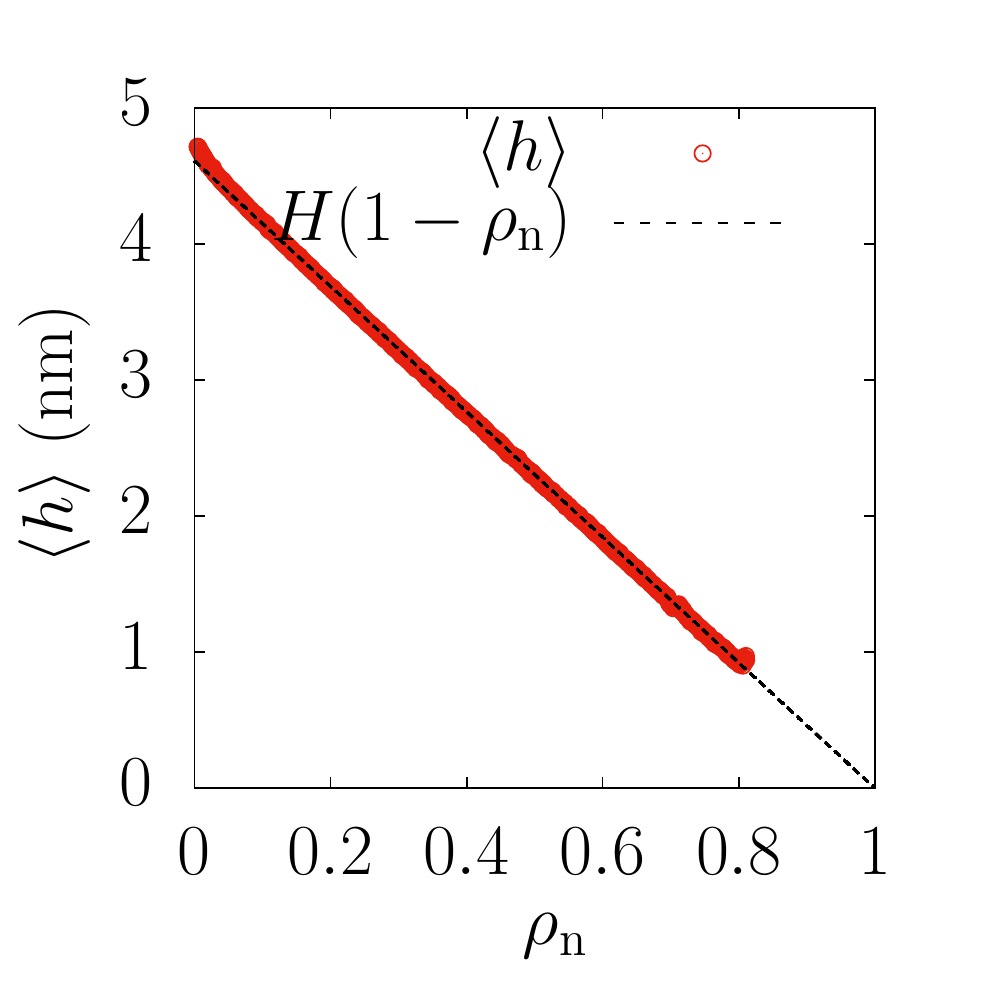}
    \caption{  
    For $\rhon<0.8$, our simulations display an intact vapor-liquid interface that is detached from, and is roughly parallel to the basal surface. 
    The average height of this vapor-liquid interface (symbols) plotted as a function of $\rhon$, agrees well with the macroscopic theory assumption of a flat interface parallel to the basal surface (dashed-line). 
    }
    \label{fig:havg}
\end{figure*}
%--------------------------------------------------------------------------------------------------------------------
%----------------------------------FIGURE 7: Pressure dependence------------------------------------------------------
\begin{figure*}[tb]
    \centering
    \includegraphics[width=0.9\textwidth]{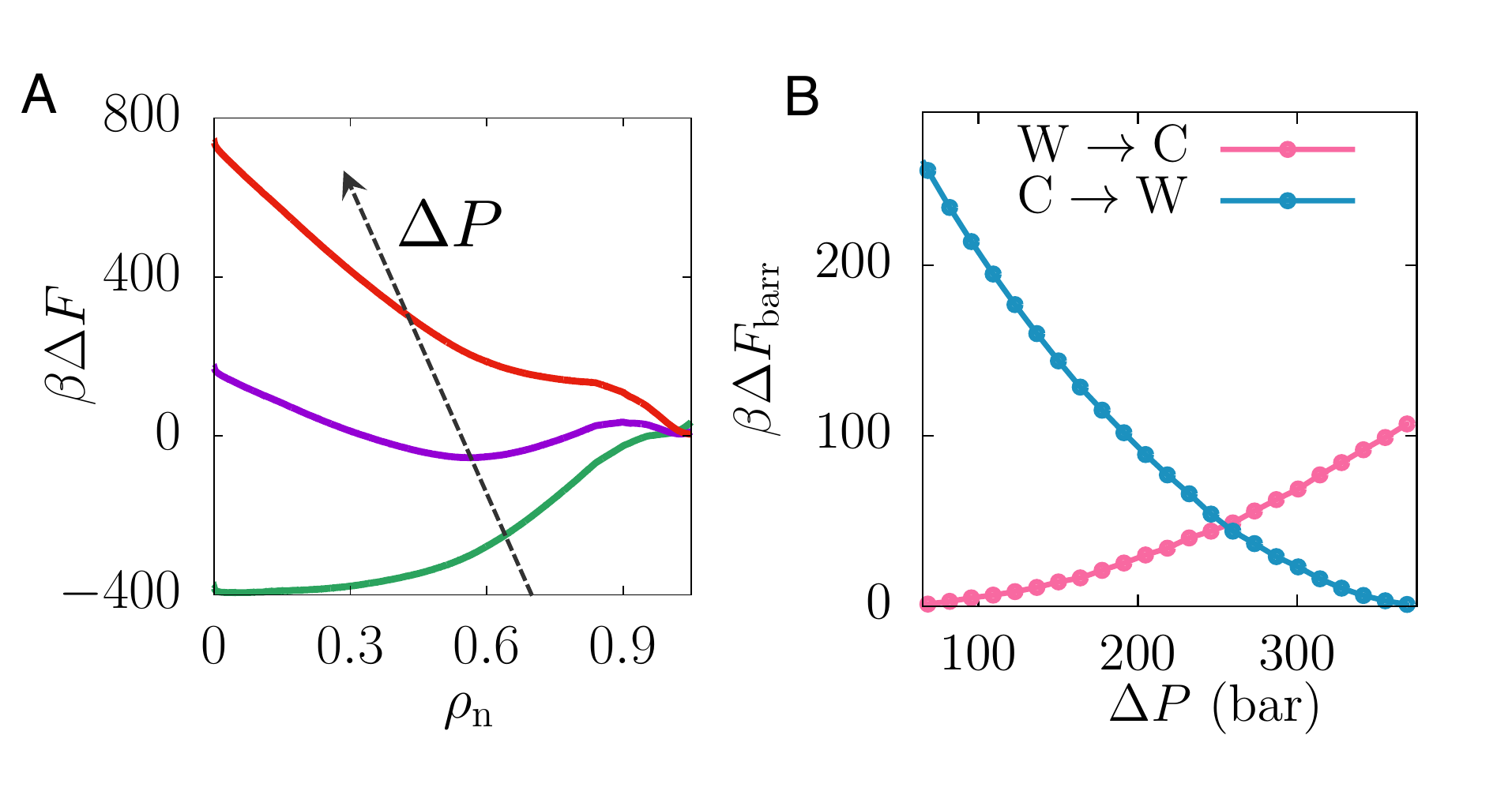}
    \caption{
    (a) $\Delta F(\rhon)$ at three different pressures for the pillared surface with a 3.5~nm nanoparticle.
    For $\Delta P=0$ (green), the Wenzel state is unstable as discussed in the main text.
    At  $\Delta P=200$~bar (purple), the Wenzel and Cassie states co-exist.
    At higher pressures, such as $\Delta P=400$~bar (red), the Wenzel state becomes the stable state.    
    (b) The barriers for the wetting and dewetting transitions are shown here as a function of $\Delta P$.
     The Cassie-to-Wenzel barrier (blue) decreases on increasing the pressure, eventually disappearing at the intrusion pressure, $\Delta P_{\rm int}=369$~bar. 
    Similarly, the Wenzel-to-Cassie barrier (magenta) barrier decreases upon decreasing pressure, and disappears at a positive extrusion pressure, $\Delta P_{\rm ext}=+68$~bar. 
     }
    \label{fig:pressureNP}
\end{figure*}
%-----------------------------------------------------------------------------------------------------------------
%
%
%----------------------------------FIGURE 8: Time series of Novel surface------------------------------------------------------
\begin{figure*}[tb]
   {\centering
    \includegraphics[width=0.9\textwidth]{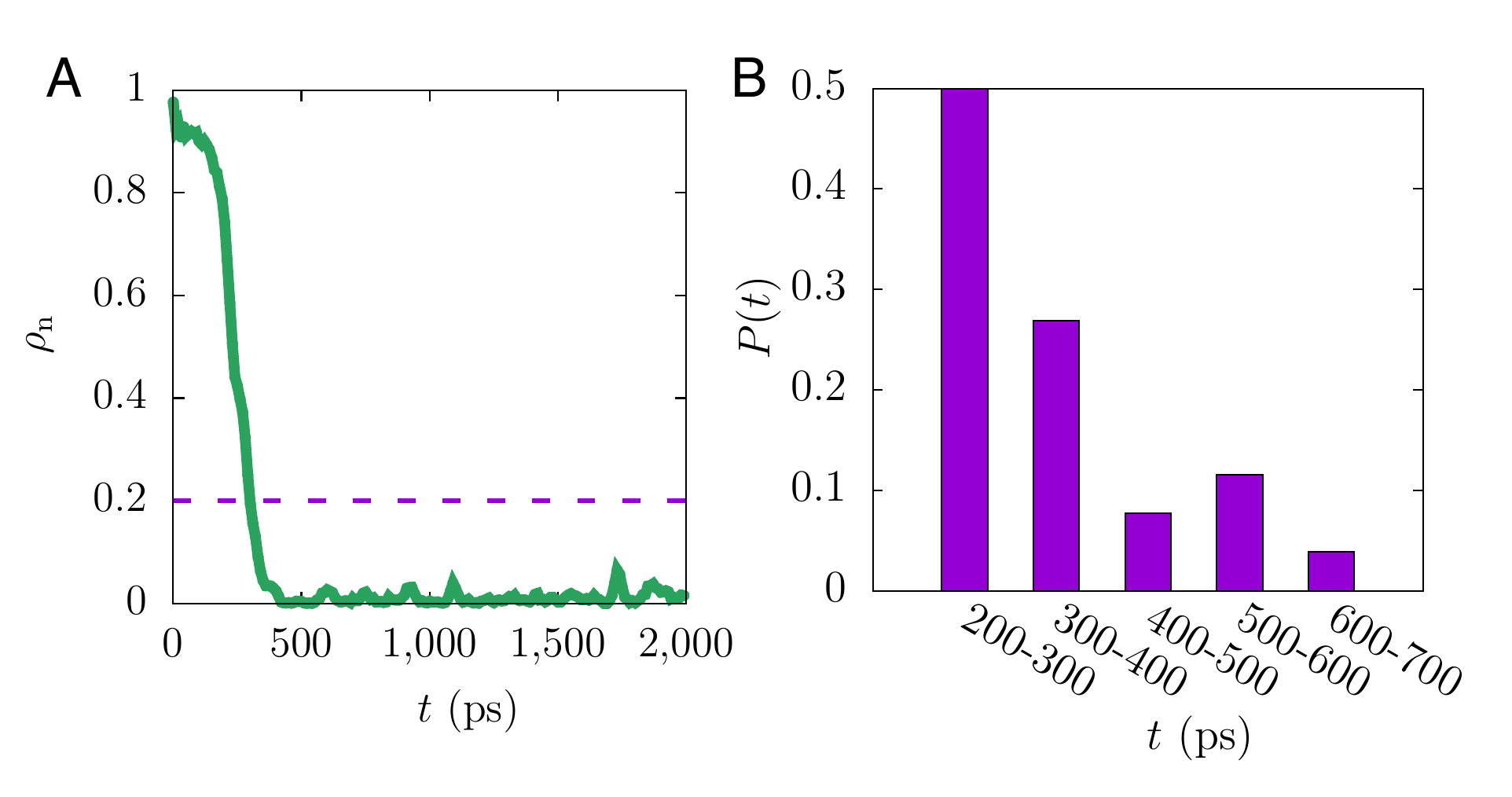}}
    \caption{ 
    (a) $\rhon$ as a function of time for a typical unbiased simulation initialized in the Wenzel state.
    (b) The distribution of times for the system to reach a configuration with $\rhon = 0.2$ was obtained from 26 trajectories initialized in the Wenzel state.
    The average time to reach $\rhon = 0.2$ is 375~ps.
       }
    \label{fig:timeNP}
\end{figure*}
%--------------------------------------------------------------------------------------------------------------------
%
%---------------------------------Figure 9: T-Free energy----------------------------------------------
\begin{figure*}[tb]
    \centering
    \includegraphics[width=0.45\textwidth]{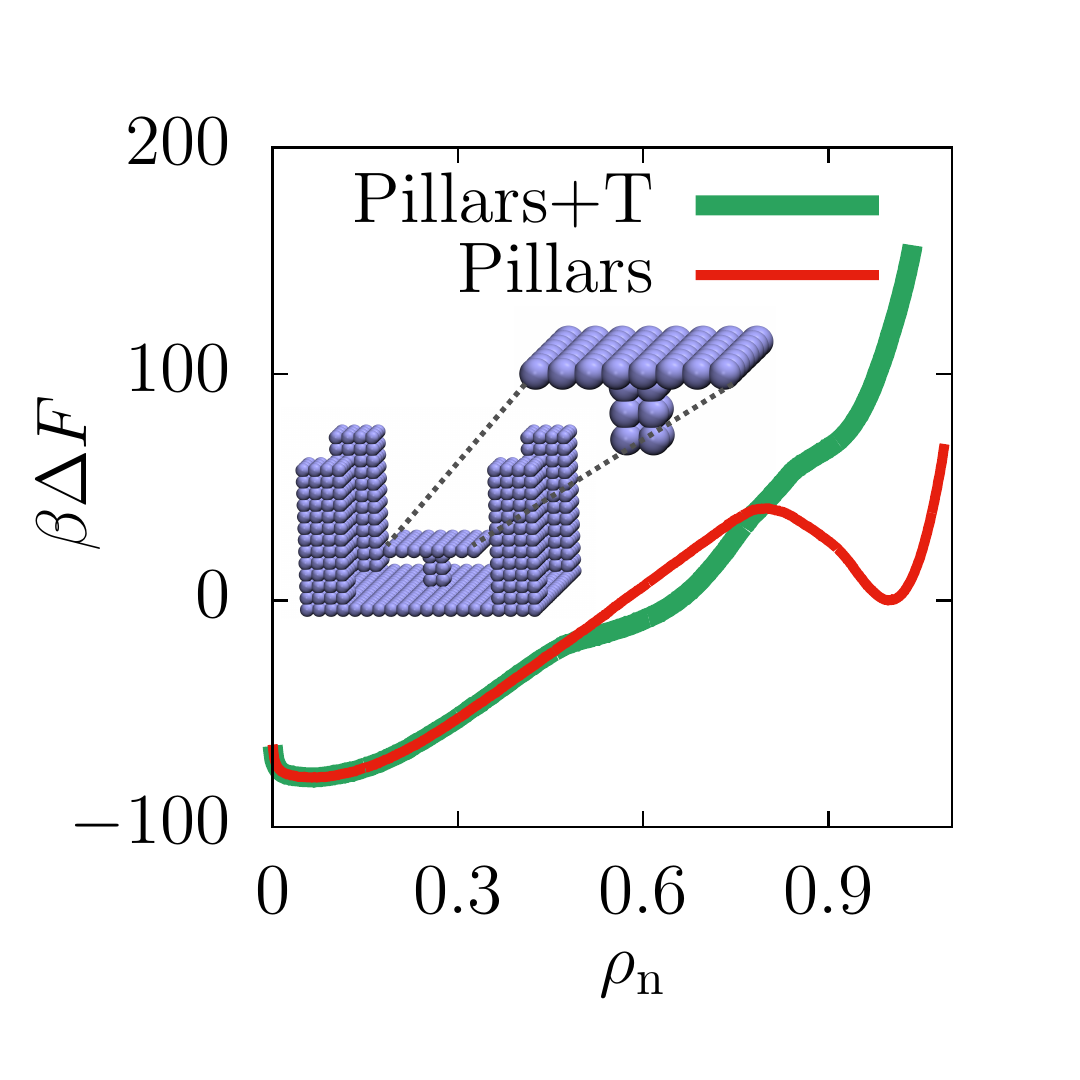}
    \caption{ Inset: Unit cell of a pillared surface with a `T' element at the center of the cell. The vertical element of the `T'  is $1~\nm$ in height and has a square cross-section with a width of $0.5~\nm$, whereas the horizontal element is composed of one layer of atoms and has a square cross-section with a width of $2~\nm$.
    The free energetics of this surface with $W=2~\nm$, $H=3~\nm$, and $S=3~\nm$, are compared to the corresponding pillared surface; the surface modification destabilizes the Wenzel state and renders it unstable.
Because the surface with the `T' element does not have a liquid basin, $N_{\rm liq}$ of the corresponding pillared surface is used to estimate $\rhon$.
	}
    \label{fig:novel}
\end{figure*}
%--------------------------------------------------------------------------------------------------------------------
%
%---------------------------------figure 10: Movie----------------------------------------------
\begin{figure*}[tb]
    \begin{center}
    \includegraphics[width=0.45\textwidth]{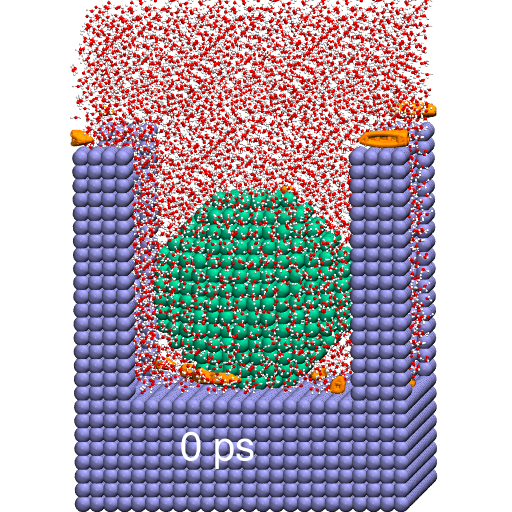}
    \end{center}
    \caption{ Movie illustrating spontaneous dewetting on the pillared surface with a 3.5~nm diameter spherical nanoparticle.
    Pillar atoms are shown as blue spheres, while the orange mesh corresponds to the instantaneous interface enveloping the vapor region.
    This snapshot corresponds to the initial configuration of the system$(t=0)$ of an unbiased simulation trajectory.
    The duration of the movie corresponds to 380~ps of simulation time.
	}
    \label{fig:movie}
\end{figure*}
%--------------------------------------------------------------------------------------------------------------------
%

\end{document}